\documentclass[namedreferences]{solarphysics}
\usepackage[hyperref,optionalrh,natbib]{spr-sola-addons} 
\usepackage{natbib}
\usepackage{graphicx}                    
\usepackage{color}                       
\usepackage{breakurl}                         
\usepackage{adjustbox}

\begin{document}

\begin{article}

\begin{opening}

\title{Quantifying the Difference Between the Flux-Tube Expansion Factor at the Source Surface and at the Alfv\'en Surface Using A Global MHD Model for the Solar Wind}

%
\author{O.~\surname{Cohen}$^{1}$\sep}

%

%
 \institute{$^{1}$ Harvard-Smithsonian Center for Astrophysics, \\60 Garden St. Cambridge, MA 02138, \\USA email: \url{ocohen@cfa.harvard.edu} }

\begin{abstract}

The potential field approximation has been providing a fast, and computationally inexpensive estimation for the solar corona's global magnetic field geometry for several decades. In contrast, more physics-based global magnetohydrodynamic (MHD) models have been used for a similar purpose, while being much more computationally expensive. Here, we investigate the difference in the field geometry between a global MHD model and the potential field source surface model (PFSSM) by tracing individual magnetic field lines in the MHD model from the Alfv\'en surface (AS), through the source surface (SS), all the way to the field line footpoint, and then back to the source surface in the PFSSM. We also compare the flux-tube expansion at two points at the SS and the AS along the same radial line. We study the effect of solar cycle variations, the order of the potential field harmonic expansion, and different magnetogram sources. We find that the flux-tube expansion factor is consistently smaller at the AS than at the SS for solar minimum and the fast solar wind, but it is consistently larger for solar maximum and the slow solar wind. We use the Wang--Sheeley--Arge (WSA) model to calculate the associated wind speed for each field line, and propagate these solar-wind speeds to 1~AU. We find a more than five hours deviation in the arrival time between the two models for 20\,\,\% of the field lines in the solar minimum case, and for 40\,\,\% of the field lines in the solar maximum case.

\end{abstract}

%

\end{opening}

\section{Introduction}
\label{Introduction}

The magnetic field above the solar photosphere and in the solar corona is weak. Therefore, it is hard to obtain observationally the three-dimensional topology of the magnetic field in the solar atmosphere. Global imaging of the solar corona (e.g., TRACE, SOHO EIT, STEREO EUV, HINODE XRT, and SDO AIA EUV/X-ray images), clearly shows magnetic loop structure in the corona. However, these images indicate the line-of-sight integral of local density and temperature structures, so they cannot provide the complete information about the magnetic field.  

While limited three-dimensional reconstruction is possible with the extensive available solar data, a complete picture of the three-dimensional structure is still limited to theoretical models. The most popular method to estimate the global three--dimensional coronal magnetic field is the so-called "Potential Field Source Surface Model" \citep[PFSSM hereafter,][]{AltschulerNewkirk69,Altschuler77}. The underlaying assumption in this method is that the large-scale structure of the solar corona does not change much over one solar rotation (the time-scale for the appearance of new active regions is longer), and it definitely does not change within the time it takes perturbations to move along the magnetic field (the Alfv\'en time-scale). With that assumption in mind, one can also assume that there are no forces in the form of electric currents applyed on the magnetic field and that the field is static. Other method to calculate more local regions in the corona allows field-aligned currents in the system. This method is called the linear/non linear "Force-free" method \citep[see review by][with references therein]{WiegelmannSakurai12}.

The PFSSM assumption that there are no currents in the system immediately implies that:

\begin{equation}
\label{Eq:1}
\nabla\times \mathbf{B}=0,
\end{equation}
so the magnetic field vector, $\mathbf{B}$, can be described as a gradient of a scalar potential, $\Psi$:

\begin{equation}
\label{Eq:2}
\mathbf{B}=-\nabla \Psi.
\end{equation}
With the additional requirement that there are no magnetic monopoles, $\nabla\cdot\mathbf{B}=0$, we can obtain the Laplace equation for the scalar potential:
\begin{equation}
\label{Eq:3}
\nabla^2\Psi=0.
\end{equation}

Equation~\ref{Eq:3} can be solved using two boundary conditions. The first boundary condition at the solar surface, $r=R_\mathrm{\odot}$, can be specified using observations of the photospheric radial magnetic field (magnetograms). The second boundary condition assumes that the coronal magnetic field becomes purely radial at some spherical distance, $R_\mathrm{ss}$, with the sphere of $r=R_\mathrm{ss}$ being known by the name "Source Surface" (SS hereafter). Once the boundary conditions are obtained, the solution for the three-dimensional structure of the coronal magnetic field can be obtained using harmonic decomposition and the Associated Legendre functions, where the harmonic coefficients in such a solution are computed using the observed photospheric field. Using well-known recursive relations for the polynomial functions and modern computers, the solution can be obtained almost instantaneously \citep[see e.g.,][]{Luhmann02}. This makes the PFSSM extremely popular to use, in contrast to expensive, physics-based models that will be discussed below.

The assumption that the SS is spherical and the assumption about its location are the main drawbacks of the method for the following reasons: First, the physical interpretation of the mathematically described SS is that it represents the surface at which all field lines are supposedly open. In reality, this surface is known as the Alfv\'en Surface (AS hereafter), which represents the collection of points along the ensemble of coronal field lines at which the solar-wind speed exceeds the local Alfv\'en speed [$u_A=B/\sqrt{4\pi\rho}$] with $B$ being the local magnetic-field strength and $\rho$ being the local mass density.  At this point, the wind's dynamic pressure overcomes the magnetic tension in the magnetic loops. As a result, the field lines are ``open`` to the heliosphere and are dragged by the radially expanding solar wind. The shape of the AS is far from spherical, and it depends on the complexity of the solar magnetic field. Even during solar minimum, where the solar magnetic field is nearly dipolar, the AS is not spherical, with the surface being much closer to the Sun near the equator (where the magnetic field is weak) than above the poles (where the magnetic field is stronger). Despite this trivial limitation of the method, very little analytical work has been done to estimate the non-hericality of the source surface in the context of the heliosphere \citep{Schulz78,Levine82,Schulz97}. 

The other problem with the outer boundary condition in the PFSSM is the assumption where to locate the SS. Traditionally, $R_\mathrm{ss}$ is set at $2.5r_\mathrm{\odot}$. This distance originated from comparison between the height of the coronal loops in the PFSSM solutions and in solar eclipse images \citep{AltschulerNewkirk69,Hoeksema84}. While this value seems to work reasonably well overall, it is more than likely that the SS can be located further out, in particular when the magnitude of the photospheric field increases. Recently, \cite{DeForest13} have used STEREO-A/COR2 data to identify returning structures in the solar corona and solar wind. These returning structures are indicators for sub-Alfv\'enic regions, meaning that these regions are below the AS. They found that overall, the AS is located at least above $6r_\mathrm{\odot}$, and it may even be located above $12r_\mathrm{\odot}$ for coronal holes and above $15r_\mathrm{\odot}$ for helmet-streamers. 

The PFSSM is commonly used as the initial condition for magnetohydrodynamic (MHD), physics-based models. The main advantage of these models over the PFSSM model is that they provide solutions for all the MHD variables of density, velocity field, magnetic field, and pressure. However, as mentioned above, they are much more computationally expensive as they require a full three-dimensional computation for all the variables. \cite{Riley06} performed a detailed comparison between MHD and PFSSM solutions for the global solar corona. They used a simplified, MHD model with a simple adiabatic energy equation \citep[constant polytropic index of $\Gamma=1.05$, and no sources:][]{MikicLinker94,Lionello98,Linker99,Mikic99} to compare solutions driven by magetograms taken during both solar minimum and solar maximum periods. They found that the coronal holes structure in the MHD and PFSSM solutions were tentatively similar, while the coronal holes' area was consistently smaller in the PFSSM solutions. This suggests that some of the magnetic flux that is being opened by the solar wind in the MHD model remains closed in the PFSSM model. One can argue that the simplified MHD model for the solar wind may not realistically energize the solar wind and that with a more realistic model developed by the same group, such as  \cite{Lionello09} or \cite{Riley11}, the discrepancies in coronal hole mapping may be even greater. Unfortunately, the MHD solution for the solar wind itself was not presented by \cite{Riley06}. 

\cite{Riley06} have also compared the shape of the SS with the shape of an iso-surface extracted from the MHD solution for nearly radial field (i.e. $|B_r|/|B|=0.97$). They found that the solar minimum configuration is consistent with analytical predictions of non-spherical PFSSM \citep{Schulz78,Levine82,Schulz97}, while the solar maximum MHD iso-surface is essentially spherical when averaging the effect of local, time-dependent active regions. We would like to stress that this radial-field criteria may be misleading in the physical context of the SS. As mentioned above, we need to consider the point where the field line is open, not where the field is radial, as a radial field can still be closed. For example, the top of the first open helmet-streamer field line can become radial at a relatively close distance from the Sun, but it is open by the solar wind at a greater distance. Indeed, the near radial sphere found by \cite{Riley06} is at much greater distance than the $R_\mathrm{ss}=2.5r_\mathrm{\odot}$ in the PFSSM. 

\cite{Lee11} compared the observed magnetic flux at 1~AU with the magnetic flux predicted by the PFSSM using different locations of the SS. They found that in order to better match the observed and predicted fluxes, the SS should be placed below $2.5r_\mathrm{\odot}$ ($1.5-1.8r_\mathrm{\odot}$). This is clearly due to the fact that the PFSSM is a static model that truncates the loops at certain height so that the lower the SS is, the more open flux is added. Some MHD models \citep[e.g.][]{Cohen07}, truncating the magnetic loops below the height at which they are actually opened by the solar wind, may over-energize the wind, leading to unrealistic solar-wind speeds in the solution. This is due to the fact that the solar-wind energization is specified according to the initial, potential-field distribution.

An important application of the PFSSM has been introduced by \cite{WangSheeley90}, where they presented empirical inverse relation between the terminal solar-wind speed, $u_\mathrm{sw}$, and amount of expansion of the magnetic-flux tube from which that particular wind originated. The expansion factor, $f_\mathrm{s}$, is defined as the ratio between the magnetic flux at the flux tube's footpoint to that at the SS:

\begin{equation}
\label{Eq:4}
f_\mathrm{s}=\frac{B(r_\mathrm{\odot})R^2_\odot}{B(R_\mathrm{SS})R^2_\mathrm{SS}},
\end{equation}
for a given magnetic-field line. The value of $f_\mathrm{s}$ is calculated from the PFSSM solution. By fitting their predicted solar-wind speed to solar wind data at 1~AU, \cite{WangSheeley90} derived the following expression for the solar-wind speed as a function of $f_\mathrm{s}$:
\begin{equation}
\label{Eq:5}
u_\mathrm{sw}=u_\mathrm{min}+A f^\mathrm{-l}_\mathrm{s}.
\end{equation}
Here $u_\mathrm{min}$ is the minimum observed solar-wind speed, $A$ is a coefficient, and $l$ is a power-law index. Since its initial release, Equation~\ref{Eq:5} has been improved to include real-time updating of the input magnetogram \citep{ArgePizzo00} and a smoothing factor that depends on the angular distance of the particular field line from the nearest coronal hole boundary, $\theta_b$. With this new factor and the real-time update, Equation~\ref{Eq:5} takes the following form:
\begin{equation}
\label{Eq:6}
u_\mathrm{sw}=u_\mathrm{min}+A\left( 1+f_\mathrm{s} \right)^\mathrm{-l} \left\{ B-C\cdot\exp{\left[ 1-(\theta_b/D)^\mathrm{m} \right]}  \right\}^\mathrm{n},
\end{equation}
where the different parameters have been continuously changed over the years to improve the fitting to solar-wind data. Equation~\ref{Eq:6} is known as the Wang--Sheeley--Arge (WSA) model which predicts the solar-wind speed at 1~AU at a given time from magnetogram data and the PFSSM \citep{McGregor11}. 

The WSA model is widely used by the heliophysics community. It has been used for years at NOAA's Space Weather Prediction Center (SWPC) and it has also been used to specify the solar-wind speed at the inner boundary of the ENLIL code \citep{Odstrcil03}: an MHD code that propagates the solar wind from the solar corona to 1~AU. The WSA-ENLIL combination has been used for many studies of the solar wind directly by individual researchers and via the Community Coordinated Modeling Center (CCMC). Despite its extensive usage, the WSA model depends heavily on the PFSSM and the determination of $f_\mathrm{s}$. It is also known to be sensitive to the magnetogram observatory and resolution as a result of different mapping of the potential field \citep[e.g.][]{Lee08,Lee11}. Recently, \cite{PoduvalZhao14} have studied the different in the expected $f_\mathrm{s}$ between PFSSM and the current sheet source surface (CSSS) model, which analytically allows field lines to change their expansion shape between $r_\mathrm{\odot}$ and $R_\mathrm{SS}$. They found that this geometrical correction improves the prediction of the solar-wind speed at 1~AU and that the WSA-ENLIL approach can be improved beyond the simple PFSSM.

In this article, we study the deviations between the PFSSM prediction for $f_\mathrm{s}$, and $f_\mathrm{s}$ calculated at the Alfv\'en point from the MHD solution for individual field lines. The MHD model accounts for the acceleration of the solar wind and the heating of the corona self-consistently. Therefore, it captures major deviations in the PFSSM field expansion and demonstrates whether or not the PFSSM wind prediction is quantitatively reasonable.

In the next section, we describe the MHD model and data that we use to obtain $f_\mathrm{s}$. We show the results in Section~\ref{Results} and discuss their implications in Section~\ref{Discussion}. We conclude our findings in Section~\ref{Conclusions}.

\section{Model and Procedure}
\label{Model}

The solar-wind MHD solution is obtained using the {BATS-R-US} generic MHD code \citep{Powell99, Toth05,Toth12}. The particular version of the code for simulating the solar corona and the solar wind includes physics-based coronal heating and acceleration of the solar wind by dissipation of the turbulent energy of Alfv\'en waves, as well as thermodynamics effects such as radiative cooling and electron heat conduction \citep{Oran13,Sokolov13,Vanderholst14}. The initial condition for the three-dimensional magnetic field is obtained using PFSSM extrapolation of magnetogram data as described above (with the SS set at $2.5r_\mathrm{\odot}$). The initial potential field becomes non-potential as the corona is heated by the dissipation of turbulent energy of Alfv\'en waves, and the solar wind is accelerated by the increased thermal and wave pressure gradient. The steady-state solution includes a hot corona and a fully developed solar wind \citep[e.g.][]{PneumanKopp71,Rusin10}, which stretches the initial magnetic field into helmet-streamer topology, which resembles the coronal structure observed during solar eclipses. 

The model uses a spherical grid that is stretched in the radial direction. The smallest grid size (at the inner boundary) is $0.01r_\mathrm{\odot}$ in the radial direction and $0.02r_\mathrm{\odot}$ in the angular direction, and it increases to $0.5r_\mathrm{\odot}$ near the outer boundary of $24r_\mathrm{\odot}$. We implement adaptive mesh refinement (AMR) near the coronal current sheet as its location dynamically changes by the wind stretching the field lines; the resolution around the current sheet is $0.1r_\mathrm{\odot}$. A convergence is typically achieved after 60,000 iterations, where this rather long time is mostly due to the detailed, second-order radiation terms that are included in the model. All of the simulations were performed on NASA's PLEIADES supercomputer at NASA AMES using 256 processors. With this setting, each solution is obtained within about six hours of physical time. 

Once a steady-state is obtained for a particular case (see Section~\ref{Data} below), we identify all of the points in the solution that belong to the AS, with a selection condition that the Alfv\'en Mach number $M_A=u_\mathrm{sw}/u_A=1\pm0.001$. Typically, we identify between 1000 to 2000 points, depending on the shape of the AS and the grid structure on it. For each identified point on the AS, we begin an Euler-based field-line tracing procedure towards the solar surface. The field-line tracing defines a spatial path step [$\pm ds_\mathrm{i}$] in the direction of the unit magnetic field vector [$\mathbf{b}_0$] for the particular point [$\mathbf{r}_\mathrm{i}$]. The magnitude of $ds_\mathrm{i}$ is the average distance between the surrounding cell centers, and the sign is determined so that the tracing is done towards, not away from the solar surface. We then map the end location of $\mathbf{r}^\prime=\mathbf{r}_\mathrm{i}+ds_\mathrm{i}\mathbf{b}_\mathrm{i}$ into the center of the nearest grid cell, $\mathbf{r}_\mathrm{i+1}$. The next step is then calculated using $\mathbf{r}^\prime=\mathbf{r}_\mathrm{i+1}+ds_\mathrm{i+1}\mathbf{b}_\mathrm{i+1}$ and so forth until the tracing crosses the solar surface. During the field-line tracing procedure, we save the following information: i) the location of the starting point on the AS, the radial distance of the starting point from the Sun, and the field magnitude at that point; and ii) the location and field magnitude at the field line footpoint on the solar surface. From each identified field-line footpoint, we begin a field-line procedure in the PFSSM solution until we reach the SS located at $r=2.5r_\mathrm{\odot}$ and we save the field magnitude at that point. All of these parameters enable us to calculate the expansion factor at the SS [$f_\mathrm{ss}$], and the expansion factor at the AS [$f_\mathrm{as}$] (with the Alfv\'en point radius [$R_\mathrm{as}$] and the field magnitude at that point [$B_\mathrm{as}$] replacing the radius and the field strength of the SS in Equation~\ref{Eq:4}).  

Once the field-line tracing procedure is complete, we have all of the information needed to compare the flux-tube expansion between the SS and the AS. 

\section{Selected Data}
\label{Data}

As the WSA model is known to be sensitive to different magnetogram input, we investigate three effects on the expansion factor differences between the SS and the AS: First, we investigate the differences between solar-minimum and solar-maximum periods using MDI data ({\sf \href{url}{www.sun.stanford.edu}}) for Carrington Rotation (CR) 1922 (for ranges April-May 1997, solar minimum) and CR 1958 (for ranges January 2000, solar maximum). Second, we study the effect of the order of the harmonic expansion [$n$] on the differences in the expansion factor, where we use the two MDI magnetograms decomposed with $n=90$ and $n=180$. Third, we compare four different magnetograms datasets for CR 2104 (October-December 2010, rising phase) observed by SDO HMI with high (3600$\times$1400) and low (720$\times$360) resolution ({\sf \href{url}{http://hmi.stanford.edu/data/synoptic.html}}), MDI, and SOLIS ({\sf \href{url}{http://solis.nso.edu}}). These four datasets are decomposed with $n=180$. 

It should be mentioned that, overall, there is a flux difference of about 20\,\% between the MDI and HMI low-resolution magnetogram, and 10\,\% between MDI and the HMI high-resolution map. The flux of the SOLIS map is significantly lower than all the other maps. Overall, all of the magnetic features in the SOLIS map are almost 80\,\% weaker than in the MDI map.

\section{Results}
\label{Results}

\begin{figure}[h!]
\centering
\noindent\includegraphics[width=4.75in]{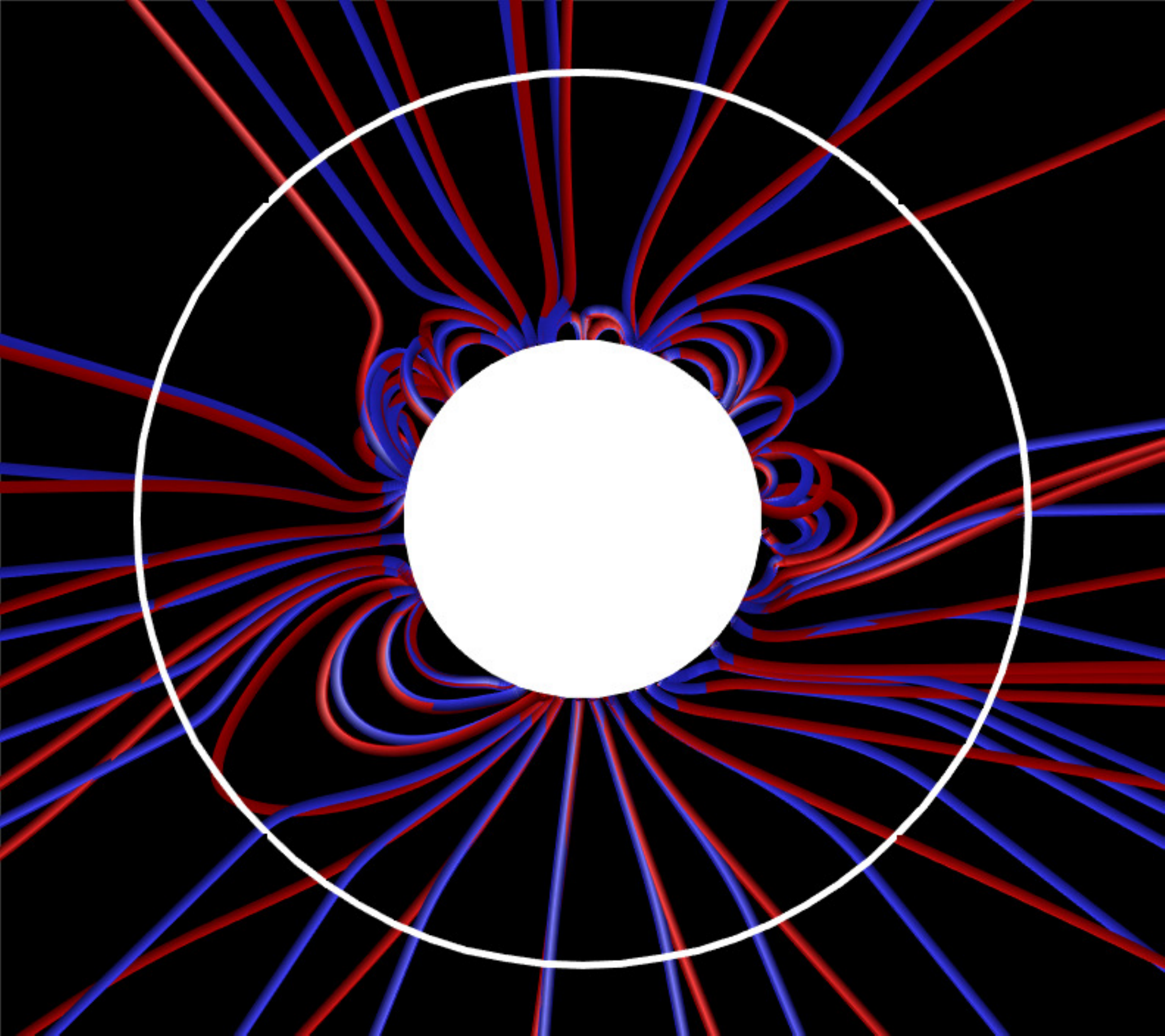}
\caption{Selected three-dimensional field lines, extracted from the same footpoint for the PFSSM (blue) and MHD solution (red) for CR 1958 MDI magnetogram with n=180. The inner sphere represents the solar surface and the solid-white line represents the SS at $r=2.5r_\mathrm{\odot}$.}
\label{fig:f1}
\end{figure}

Figure~\ref{fig:f1} demonstrates the similarity of the three-dimensional field geometry between the PFSSM and the MHD model below the SS. Each pair of three-dimensional field lines are selected at the same footpoint, and are then drawn in each of the three-dimensional solutions. It can be seen that the geometry of most of the field lines, which are not associated with closed loops is very similar in both models below the SS. The main notable differences are for the field lines that belong to closed loops or overlaying closed loops. This is due to the fact that the solar wind existing in the MHD solution stretches and opens some of the closed loops in the PFSSM. 

\begin{figure}[h!]
\centering
\noindent\includegraphics[width=3.in]{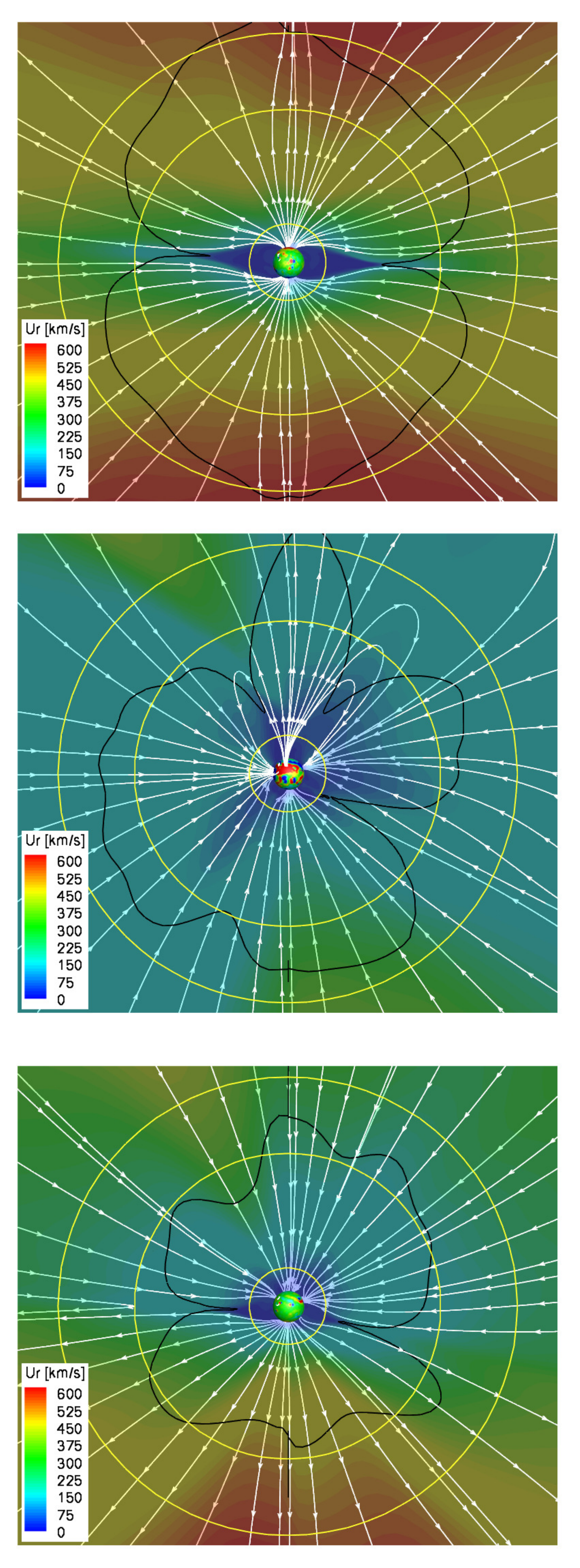}
\caption{Three meridional slices colored with solar-wind speed contours for MDI CR 1922 (left), MDI CR 1958 (middle), and HMI high-resolution CR 2104 (right). Selected three-dimensional field lines (not two-dimensional projection) are shown in white, and the AS is shown as solid-black line. The three yellow circles represent distances of $2.5$, $10$, and $15r_\mathrm{\odot}$, respectively. The size of the displayed box is $17.5r_\mathrm{\odot}$ in each direction.}
\label{fig:f2}
\end{figure}

Figure~\ref{fig:f2} shows meridional slices of the MHD solutions using MDI magnetograms for CR 1922 and CR 1958, and an HMI high-resolution magnetogram for CR 2104. The slices are colored with solar-wind speed contours, along with the AS indicated as black line. The dipolar topology of solar minimum period (CR 1922) is seen in the solar-wind structure, along with a thin, equatorial region of slow solar wind near the current sheet, and a rather symmetric AS. During the solar maximum period (CR 1958), the field topology is more complex, leading to an extensive slow-wind region and a tilted current sheet. The field complexity is also reflected in the shape of the AS. CR 2104 (November-december 2010) occurred at the beginning of the rising phase after the extended solar minima at the end of Solar Cycle 23. The dipolar topology and the aligned current sheet are still present in the solution, while the region occupied by slow wind is more extended than in the solution for CR 1922. The shape of the AS represents an intermediate state between solar minimum and solar maximum. For all solutions, the distance of the AS is in an overall agreement with \cite{DeForest13}. 

\begin{figure}[h!]
\centering
\noindent\includegraphics[width=4.5in]{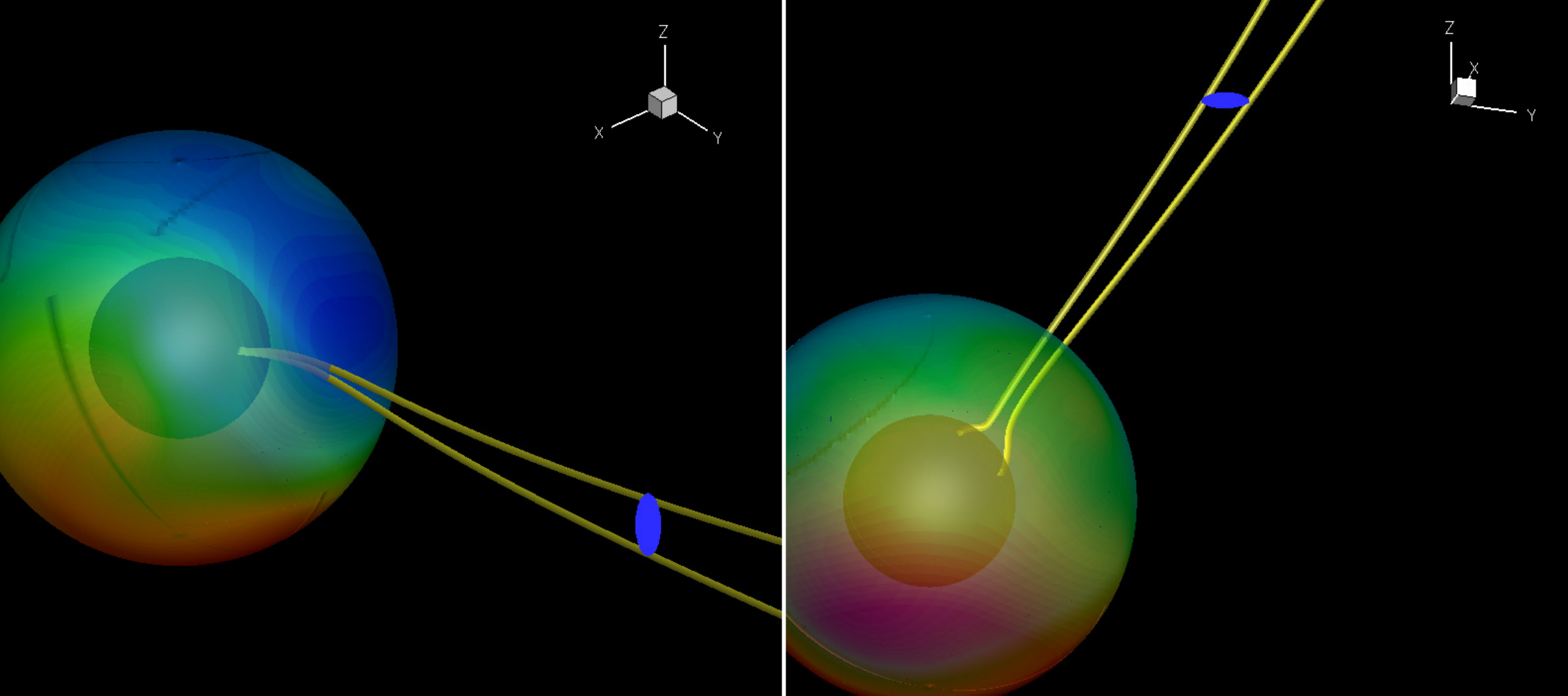}
\caption{Two selected flux tubes with their boundaries shown in yellow, and their expansion from the solar surface to the AS. The inner sphere represents the solar surface, the outer, transparent sphere represents the SS at $r=2.5r_\mathrm{\odot}$ (both spheres are colored with the radial field strength), and the blue ellipse represents the area of the flux tube at the AS.}
\label{fig:f3}
\end{figure}

Figure~\ref{fig:f3} shows how the flux-tube expansion changes from the solar surface to the AS for two selected flux tubes. One flux tube (left) gradually expands with rather constant rate. This topology is typical of coronal holes and fast-wind regions. The flux tube on the right represents open flux tube at the boundary of a closed loop, which is a typical source region for the slow solar wind. Here the expansion significantly changes from the lower regions to the AS. Figure~\ref{fig:f3} demonstrates the physical essence of the relation between the solar-wind speed and $f_\mathrm{s}$ shown in Equation~\ref{Eq:5} and Equation~\ref{Eq:6}. 

\subsection{Tracing the Field Lines in the MHD Model and the PFSSM from the Same Footpoint}
\label{sec:MDHPFSSM}

\begin{figure}[h!]
\centering
\noindent\includegraphics[width=4.75in]{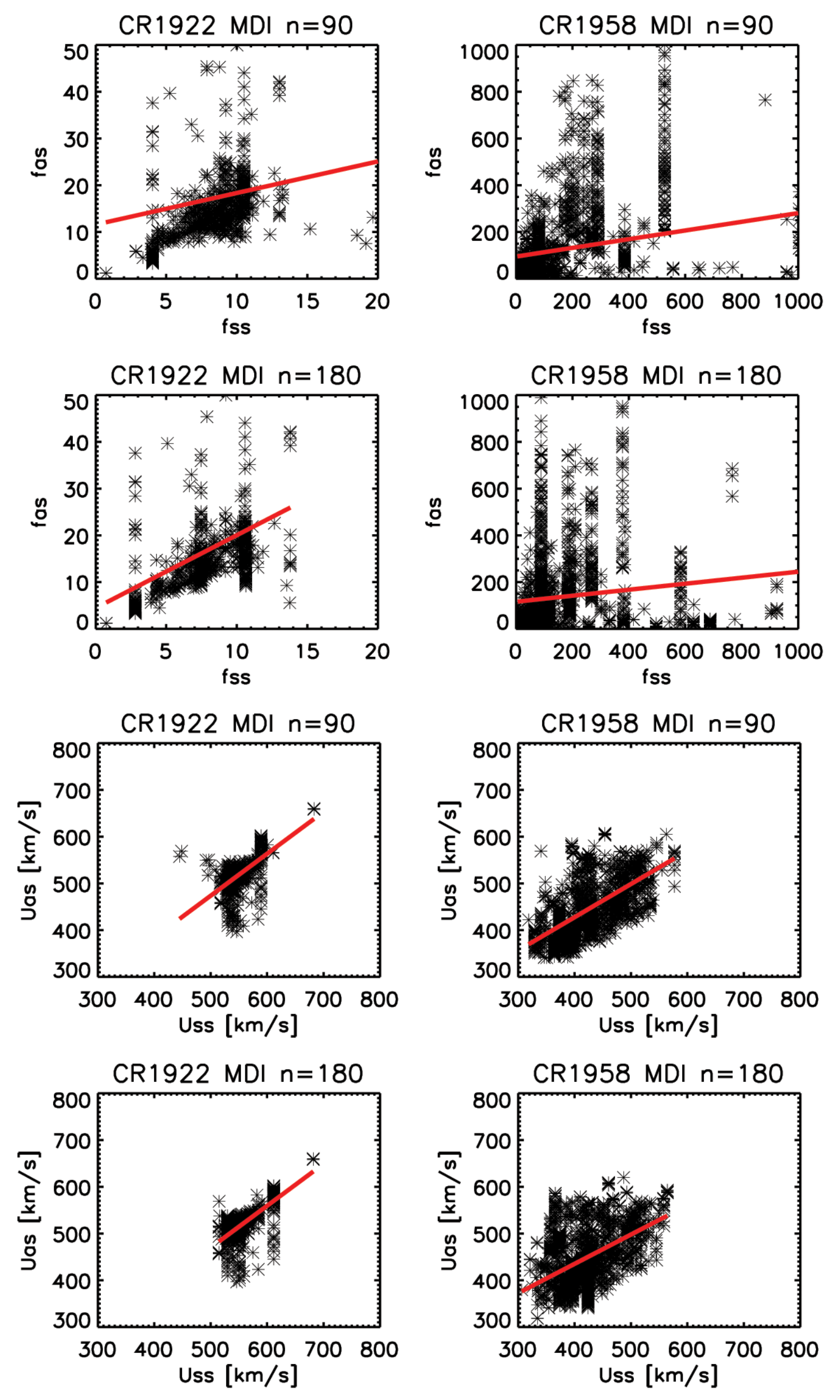}
\caption{Scatter plots showing the value of the expansion factor at the SS and at the AS, based on field line tracing, for MDI magnetograms taken during CR 1922  and CR 1958 using n=90 and n=180. Also shown is a comparison between the wind speed associated with each value of the expansion factor based on Equation~\ref{Eq:7}. Each panel includes a linear fit marked as solid red line. The parameters of the linear fit for each case are summarized in Table~\ref{Table:t1}.} 
\label{fig:f4}
\end{figure}

\begin{figure}[h!]
\centering
\noindent\includegraphics[width=4.75in]{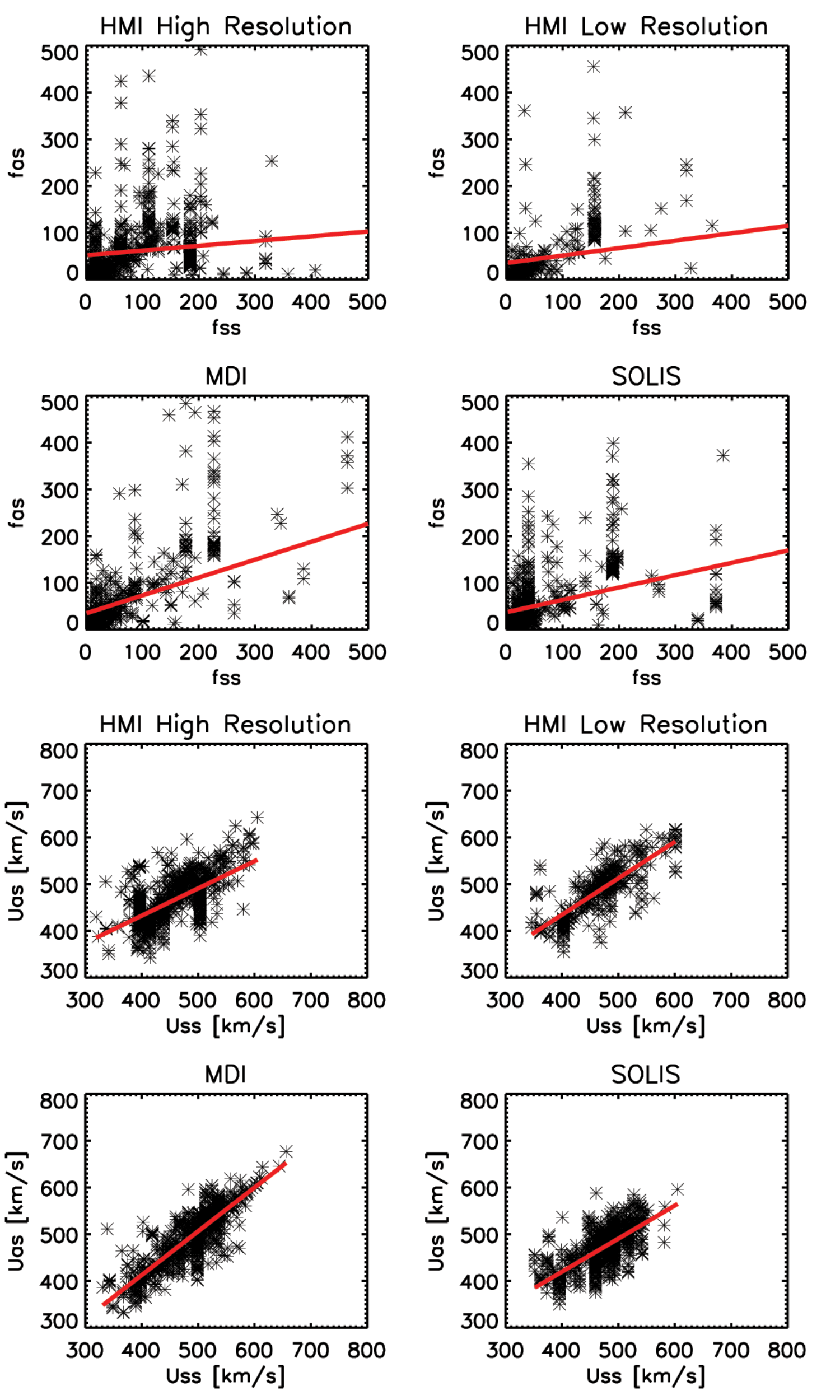}
\caption{Scatter plots showing the value of the expansion factor at the SS and at the AS, based on field line tracing, for magnetograms taken during CR 2104. Also shown is a comparison between the wind speed associated with each value of the expansion factor based on Equation~\ref{Eq:7}. Each panel includes a linear fit marked as solid red line. The parameters of the linear fit for each case are summarized in Table~\ref{Table:t1}.}
\label{fig:f5}
\end{figure}

The results of the field-line tracing are shown in Figure~\ref{fig:f4} and Figure~\ref{fig:f5}. Each point in the top four scatter plots represents the value of the expansion factor at the SS [$f_\mathrm{ss}$] {\it vs.} the value of the expansion factor at the AS [$f_\mathrm{as}$]. Figure~\ref{fig:f4} shows the results for MDI magnetogram data taken during CR 1922 and CR 1958 with n=90 and n=180. Figure~\ref{fig:f5} shows the results for magnetogram data taken by the four magnetogram sources for CR 2104. The bottom four scatter plots in each figure show the wind speed associated with a given value of the expansion factor. The associated speed is calculated using the relation:

\begin{equation}
\label{Eq:7}
u_\mathrm{sw}=\left[240+ \frac{500}{(1+f_\mathrm{s})^{0.22}} \right]\;\mathrm{km\;s^{-1}}.
\end{equation}
There are a number of different coefficients in the literature that were modified to tune the model prediction of the solar-wind speed at 1AU. Here, we choose the average coefficients between the different published values \citep[e.g.,][the value of $A=500\;km\;s^{-1}$ is also a representative mid-point between the fast and the slow solar wind]{ArgePizzo00,McGregor11}. In order to isolate the effect of the expansion factor, we also neglect the smoothing factor in Equation~\ref{Eq:6} that depends on the distance from the coronal hole boundary as it is important mostly for real-time tracking of the solar wind from the Sun to the Earth.

For each scatter plot in Figure~\ref{fig:f4} and Figure~\ref{fig:f5}, we calculate a linear fit in the form of $f_\mathrm{as}=\alpha+\beta f_\mathrm{ss}$, and $u_\mathrm{as}=\gamma+\delta u_\mathrm{ss}$. We also calculate the errors of these coefficients, as well as the correlation parameters, and all of these parameters are summarized in Table~\ref{Table:t1}. Overall, the correlation for the expansion factors ranges between 0.1--0.4, with the correlation being notably smaller for the MDI case of CR 1922 with n=90. The correlation for $u$ ranges between 0.44--0.86. The errors of the slope $\beta$ for the MDI cases are higher than the cases for CR 2104, with the exception of CR 1958 with n=90. 

Overall, the the correlation for the expansion factor between the two models is poor and the scatter of the plots is quite large, even for the "simple" solar minimum configuration. The correlation for the associated wind speeds is better, but it is still less than 0.7 for most of the magnetogram sources. Since the expansion factor represents the field geometry, this low correlation is another indication that the three-dimensional field geometry in the MHD solution is significantly different than that of the PFSSM. The parameters presented in Table~\ref{Table:t1} provide scaling parameters between the two models for selected cases of magnetogram sources. More extensive calculations of longer periods and for all magnetogram sources are left for future work.

\begin{table}
\caption{Linear fit and correlation parameters for field line tracing from the SS to the AS}
\begin{adjustbox}{max width=4.5in}
\begin{tabular}{ccccccc}
\hline
Magnetogram  & $\alpha$ & $\beta$ & $f_\mathrm{s}$ & $\gamma$ & $\delta$ & $u$ \\
Input &  &  & Correlation &  & & Correlation \\
\hline
MDI CR 1922 n=90 & 12$\pm$2 & 0.7$\pm$0.19 & 0.1 & 25$\pm$24 & 0.89$\pm$0.04 & 0.6\\
MDI CR 1922 n=180 & 5$\pm$2 & 1.6$\pm$0.20 & 0.3 & 22$0\pm$17 & 0.89$\pm$0.03 & 0.7\\
MDI CR 1958 n=90 & 95$\pm$5 & 0.2$\pm$0.01 & 0.4 & 143$\pm$10 & 0.71$\pm$0.02 & 0.6\\
MDI CR 1958 n=180 & 116$\pm$6 & 0.1$\pm$0.20 & 0.20 & 184$\pm$13 & 0.63$\pm$0.03 & 0.4\\
HMI high-res CR 2104 & 51$\pm$3 & 0.1$\pm$0.01 & 0.3 & 199$\pm$10 & 0.58$\pm$0.02 & 0.6\\
HMI low-res CR 2104 & 35$\pm$4 & 0.2$\pm$0.02 & 0.4 & 122$\pm$13 & 0.78$\pm$0.03 & 0.8\\
MDI CR 2104               & 35$\pm$5 & 0.4$\pm$0.02 & 0.5 & 38$\pm$9 & 0.93$\pm$0.02 & 0.9 \\
SOLIS CR 2104          & 37$\pm$2 & 0.3$\pm$0.02 & 0.4 & 136$\pm$11 & 0.71$\pm$0.02 & 0.7 \\
\hline
\end{tabular}
\end{adjustbox}
\label{Table:t1}
\end{table}

Table~\ref{Table:t2} shows the percentage of field lines for which $f_\mathrm{as}$ is greater than $f_\mathrm{ss}$ and vica versa for all cases. Overall, $f_\mathrm{as}$ is greater than $f_\mathrm{ss}$ for solar-minimum periods, and $f_\mathrm{ss}$ is greater than $f_\mathrm{as}$ for solar-maximum periods. The latter trend appears also for CR 2104 (rising phase of Solar Cycle 24).

\begin{table}
\caption{Percentage of the magnitudes of $f_\mathrm{as}$ vs. $f_\mathrm{ss}$ for field line tracing from the SS to the AS}
\begin{tabular}{ccc}
\hline
Magnetogram Input  & $f_\mathrm{as}>f_\mathrm{ss}$ & $f_\mathrm{ss}>f_\mathrm{as}$   \\
\hline
MDI CR 1922 n=90 & 95\,\% & 5\,\% \\
MDI CR 1922 n=180 & 97.5\,\% & 2.5\,\% \\
MDI CR 1958 n=90 & 36\,\% & 64\,\% \\
MDI CR 1958 n=180 & 40\,\% &60\,\% \\
HMI high-res CR 2104 & 35\,\% & 65\,\% \\
HMI low-res CR 2104 & 35\,\% & 65\,\% \\
MDI CR 2104 & 18\,\% & 82\,\% \\
SOLIS CR 2104 & 50\,\% & 50\,\%\\
\hline
\end{tabular}
\label{Table:t2}
\end{table}

\subsection{Assuming a Purely Radial Field Above the Source Surface}
\label{sec:MDHPFSSMAngle}

The WSA model commonly produces a spherical distribution of the expansion factor and the associated wind speed at the SS (i.e., $f_\mathrm{s}(R_\mathrm{SS},\theta,\phi)$ and $u_\mathrm{sw}(R_\mathrm{SS},\theta,\phi)$). The distribution of $u_\mathrm{sw}$ is then propagated ballistically to predict the wind speed distribution at 1~AU, $u_\mathrm{sw}(r=1AU,\theta,\phi)$, assuming that the field is purely radial above the SS. We can take advantage of the data calculated here to test this approach and compare it with the tracing of field lines in both models from the same footpoint. Here, instead of starting our tracing at the AS, we start at a given point on the SS, and trace it back to the Sun in order to find its footpoint. This tracing identifies "open" field lines in the PFSSM. Instead of tracing the field lines between the MHD model and the PFSSM model, we now simply find the point on the AS with the same angle as the point on the SS ($(r_\mathrm{as},\theta_i,\phi_i)$ and $(r_\mathrm{ss},\theta_i,\phi_i)$). Once the three points are identified, we calculate the expansion factor in the same manner as in the previous section. 

\begin{figure}[h!]
\centering
\noindent\includegraphics[width=4.75in]{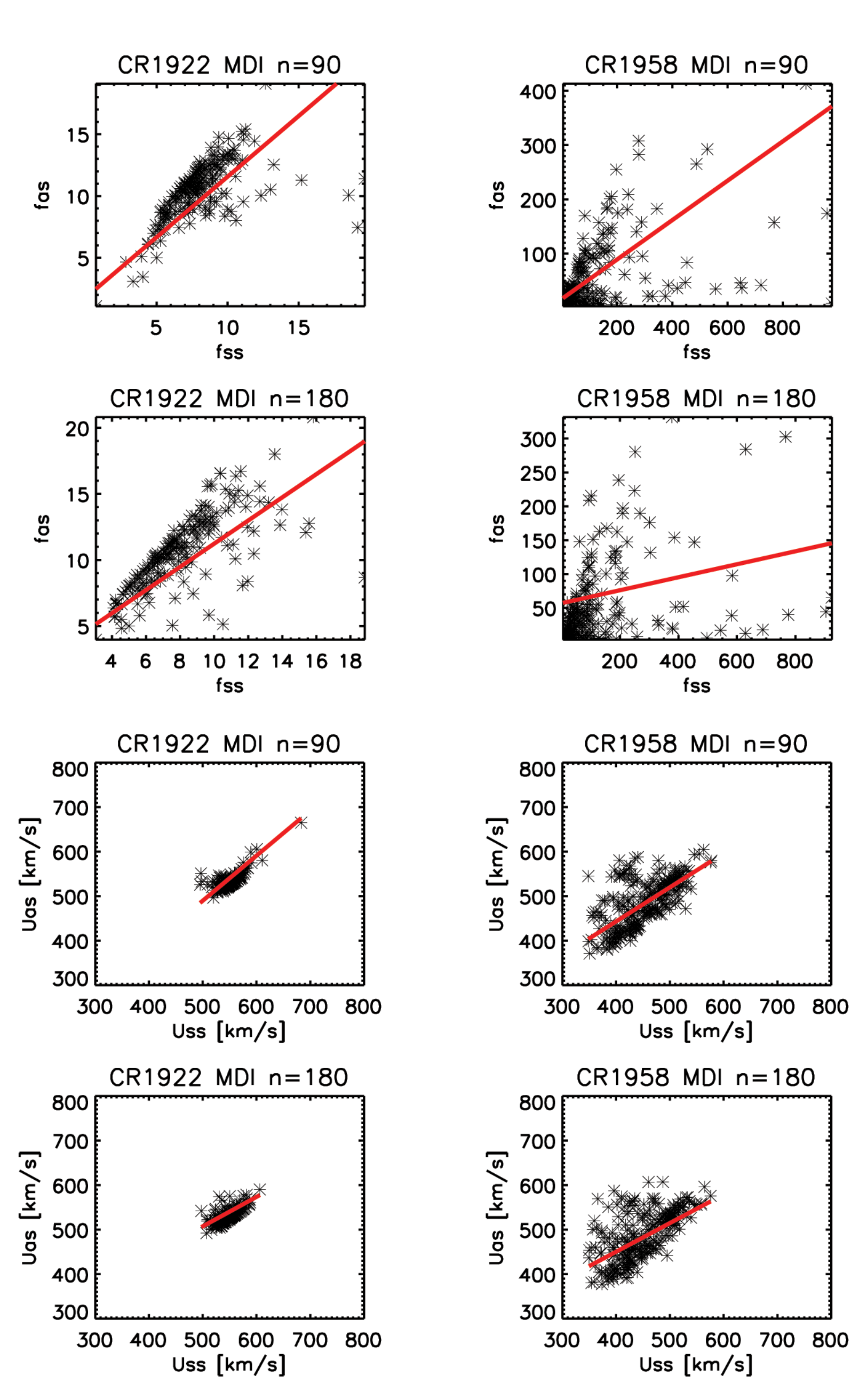}
\caption{Scatter plots showing the value of the expansion factor at the SS and at the AS, based on radial propagation, for MDI magnetograms taken during CR 1922  and CR 1958 using n=90 and n=180. Also shown is a comparison between the wind speed associated with each value of the expansion factor based on Equation~\ref{Eq:7}. Each panel includes a linear fit marked as solid red line. The parameters of the linear fit for each case are summarized in Table~\ref{Table:t3}.}
\label{fig:f6}
\end{figure}

\begin{figure}[h!]
\centering
\noindent\includegraphics[width=4.75in]{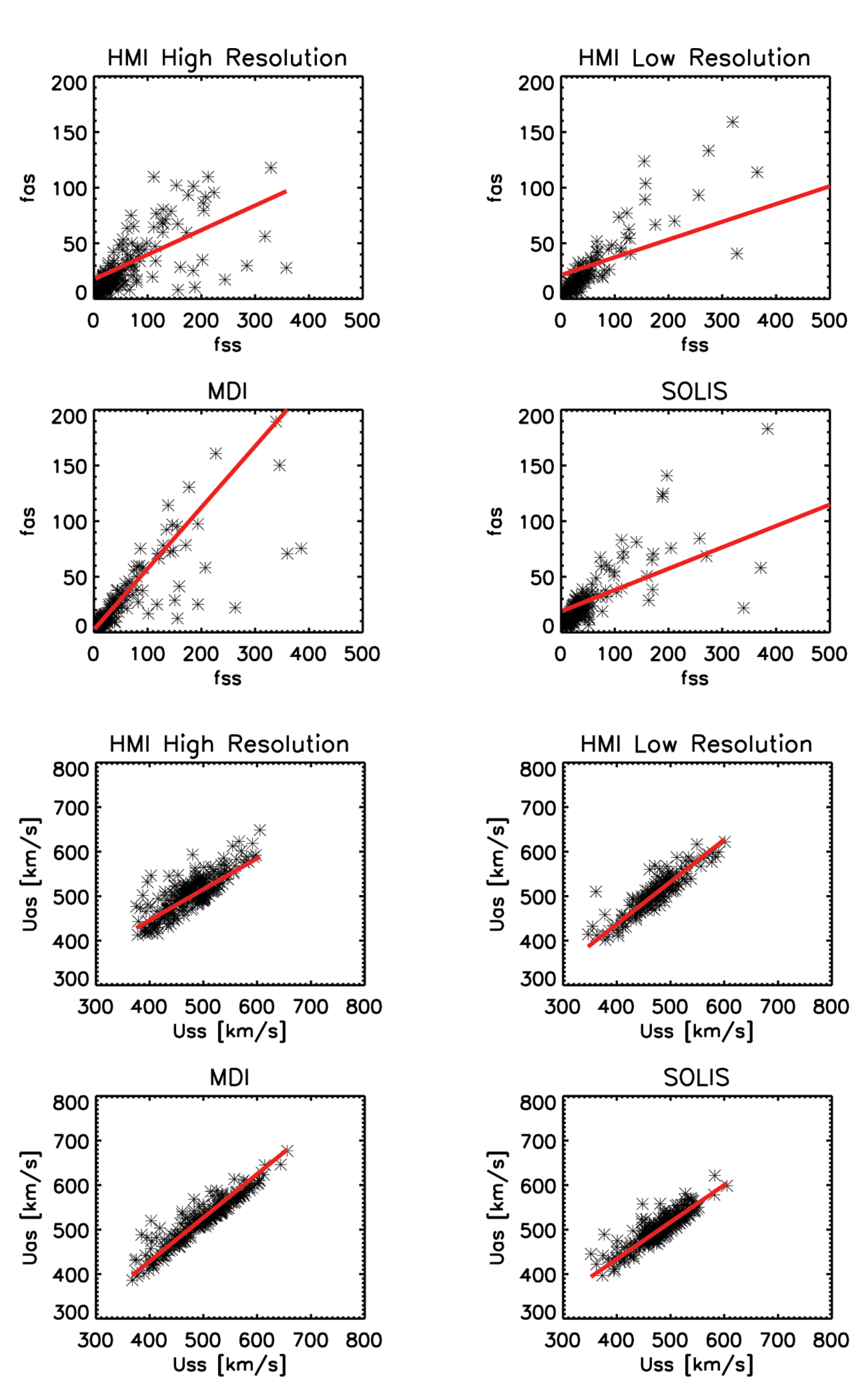}
\caption{Scatter plots showing the value of the expansion factor at the SS and at the AS, based on radial propagation, for magnetograms taken during CR 2104. Also shown is a comparison between the wind speed associated with each value of the expansion factor based on Equation~\ref{Eq:7}. Each panel includes a linear fit marked as solid red line. The parameters of the linear fit for each case are summarized in Table~\ref{Table:t3}.}
\label{fig:f7}
\end{figure}

Figure~\ref{fig:f6} and Figure~\ref{fig:f7} show the result for the this radial propagation from the SS to the AS. A notable feature is that these results are less scattered than the results for field line tracing between the models. Tables~\ref{Table:t3} and \ref{Table:t4} show the properties of the linear fit and correlation functions for the radial propagation. Overall, the correlation between the values of $f_\mathrm{s}$ at the SS and the AS are better than the full field tracing, and almost all the correlations for $f_\mathrm{s}$ are about 0.5 or higher (with the exception of CR 1958 MDI with n=180), better than the values presented in Table~\ref{Table:t1}. Table~\ref{Table:t4} shows the percentage of cases at which $f_\mathrm{as}$ is larger than $f_\mathrm{ss}$ and visa versa. It is consistent with the values appear in Table~\ref{Table:t2} except that here, $f_\mathrm{ss}>f_\mathrm{as}$ in almost all the points in the cases for CR 2104.

The result of these calculations could be used to get better predictions from the common WSA approach based on PFSSM only.

\begin{table}
\caption{Linear fit and correlation parameters for radial propagation from the SS to the AS}
\begin{adjustbox}{max width=4.5in}
\begin{tabular}{ccccccc}
\hline
Magnetogram  & $\alpha$ & $\beta$ & $f_\mathrm{s}$ & $\gamma$ & $\delta$ & $u$ \\
Input &  &  & Correlation &  & & Correlation \\
\hline
MDI CR 1922 n=90 & 2.1$\pm$0.2 & 0.94$\pm$0.03 & 0.76 & -10$\pm$11 & 1.00$\pm$0.02 & 0.87\\
MDI CR 1922 n=180 & 2.8$\pm$0.3 & 0.83$\pm$0.03 & 0.64 & 181$\pm$17 & 0.65$\pm$0.03 & 0.59\\
MDI CR 1958 n=90 & 48.9$\pm$2.6 & 0.13$\pm$0.01 & 0.48 & 141$\pm$8 & 0.75$\pm$0.02 & 0.69\\
MDI CR 1958 n=180 & 63.4$\pm$2.3 & 0.10$\pm$0.01 & 0.29 & 202$\pm$12 & 0.62$\pm$0.03 & 0.46\\
HMI high-res CR 2104 & 1.0$\pm$0.1 & 0.82$\pm$0.01 & 0.53 & 69.8$\pm$4.7 & 0.88$\pm$0.01 & 0.80\\
HMI low-res CR 2104 & 20.0$\pm$1.7 & 0.18$\pm$0.01 & 0.65 & 56.7$\pm$7.8 & 0.95$\pm$0.02 & 0.94\\
MDI CR 2104               & 2.0$\pm$0.1 & 0.55$\pm$0.01 & 0.91 & 212.5$\pm$8.9 & 0.63$\pm$0.02 & 0.94 \\
SOLIS CR 2104          & 19.0$\pm$1.0 & 0.19$\pm$0.01 & 0.63 & 101.2$\pm$6.6 & 0.83$\pm$0.12 & 0.89 \\
\hline
\end{tabular}
\end{adjustbox}
\label{Table:t3}
\end{table}

\begin{table}
\caption{Percentage of the magnitudes of $f_\mathrm{as}$ vs. $f_\mathrm{ss}$ for radial propagation from the SS to the AS}
\begin{tabular}{ccc}
\hline
Magnetogram Input  & $f_\mathrm{as}>f_\mathrm{ss}$ & $f_\mathrm{ss}>f_\mathrm{as}$   \\
\hline
MDI CR 1922 n=90 & 82\,\% & 18\,\% \\
MDI CR 1922 n=180 & 77\,\% & 23\,\% \\
MDI CR 1958 n=90 & 13\,\% & 87\,\% \\
MDI CR 1958 n=180 & 25\,\% &75\,\% \\
HMI high-res CR 2104 & 6\,\% & 94\,\% \\
HMI low-res CR 2104 & 1\,\% & 99\,\% \\
MDI CR 2104 & 1\,\% & 99\,\%\\
SOLIS CR 2104 & 5\,\% & 95\,\%\\
\hline
\end{tabular}
\label{Table:t4}
\end{table}

\section{Discussion}
\label{Discussion}

The ultimate goal of the investigation presented here is to quantify the change in the predicted solar-wind speed as a function of the expansion factor when the latter is calculated at the SS and at the AS. Our results show consistent deviation between $f_\mathrm{ss}$ and $f_\mathrm{as}$, ranging from 10--50\,\% difference and even more. 

\subsection{The Effect of Solar Cycle Variations and the Order of the Harmonic Expansion}
\label{SolarCycleOrder}

Figure~\ref{fig:f4} shows that the the results for CR 1922 (solar minimum) represent mostly field lines with very small expansion and associated fast solar-wind speed above $500\;km\;s^{-1}$, and a rather small scatter around the linear-fit line. This is a good representation of the rather stable observed fast solar wind \citep{McComas07}. For the vast majority of these field lines (i.e. for the fast solar wind), the PFSSM under-estimates the expansion factor at the Alfv\'en surface. This is due to the fact that while the expansion is small in these nearly parallel flux tubes, their geometry is changed by the solar wind beyond the source surface and some further expansion occurs. 

In contrast, the results for CR 1958 (solar maximum) show that the majority of the field lines have large expansion factor (above 100) , slow associated solar-wind speed, and a large scatter around the linear-fit line. This is a good representation of the sporadic observed slow solar wind \citep{McComas07}. The majority of these field lines show $f_\mathrm{ss}$ greater than $f_\mathrm{as}$, which means that for the slow solar wind, the PFSSM over-estimates the expansion factor. This is the result of the artificial truncation and opening of the closed loops at the SS, which forces the field lines to be radial at the SS. In the more realistic, MHD solution, these field lines are not necessarily radial at the SS, and the result is an expansion factor that is consistently larger in the PFSSM comparing to the MHD model for large  values of $f_\mathrm{s}$.

Interestingly, CR 2104, which represents the rising phase, shows a trend similar to the solar-maximum case, as it already introduces significant amount of field lines associated with slow solar wind. Both of the trends mentioned above provide an indication that the solar wind continues to accelerate and reshape the coronal magnetic field beyond the SS. It has been suggested by \cite{Gilbert07} that instead of imposing the SS at a particular location, the field needs to relax to magnetic-pressure equilibrium in order to be fully radial.

All the trends mentioned above also appear in the comparison between the values at the SS and AS for the same angle (radial propagation). Here as well, $f_\mathrm{s}(SS)$ over-estimates $f_\mathrm{s}(AS)$ for solar maximum and the rising phase. From these results we conclude that the expansion factor is similar at the SS and the AS, only during pure solar minimum conditions, when a significant amount of solar wind comes from coronal holes (fast wind). The flux tubes associated with this kind of wind are nearly parallel to each other, and the assumption of a pure radial field that drops with $r^2$ ($f_\mathrm{s}\approx 1$) is valid. For other times, when most of the wind comes from the regions of streamers/pseudo-streamers, the field is far from radial above the SS. 

It seems like the radial propagation smears up the field lines at which the non-radial drop in the field is the greatest. As a result, the scatter around the linear fit is smaller, and the correlation is better than the full field line tracing. 

The order of the spherical-harmonic expansion shows very little difference in the overall MDI solutions. Both the linear fit and the comparison between the magnitudes of $f_\mathrm{ss}$ and $f_\mathrm{as}$ are quantitatively similar. This means that despite the difference in the field geometry due to the different resolution, the overall mapping of the field lines in the PFSSM comparied to the MHD model is consistent for each line, regardless of the number of field lines (i.,e. the magnetogram resolution).

\subsection{The Effect of Magnetogram Source}
\label{MagnetogramSource}

The solutions for CR 2104 show very little difference between the magnetogram sources. This is quite consistent with the result above that the solution is not sensitive to the order of the harmonic expansion. The solutions for the HMI magneto grams are almost identical, while the MDI and the SOLIS solution show similar trends, but with slightly different parameters. The most notable difference appears in the comparison between the magnitude of $f_\mathrm{ss}$ and $f_\mathrm{as}$ for the SOLIS magnetogram solution, where one is equally bigger than the other. This is probably due to the fact that the overall magnetic flux and the magnitude of the strong-field features is lower than the other magnetograms.

\begin{figure}[h!]
\centering
\noindent\includegraphics[width=4.75in]{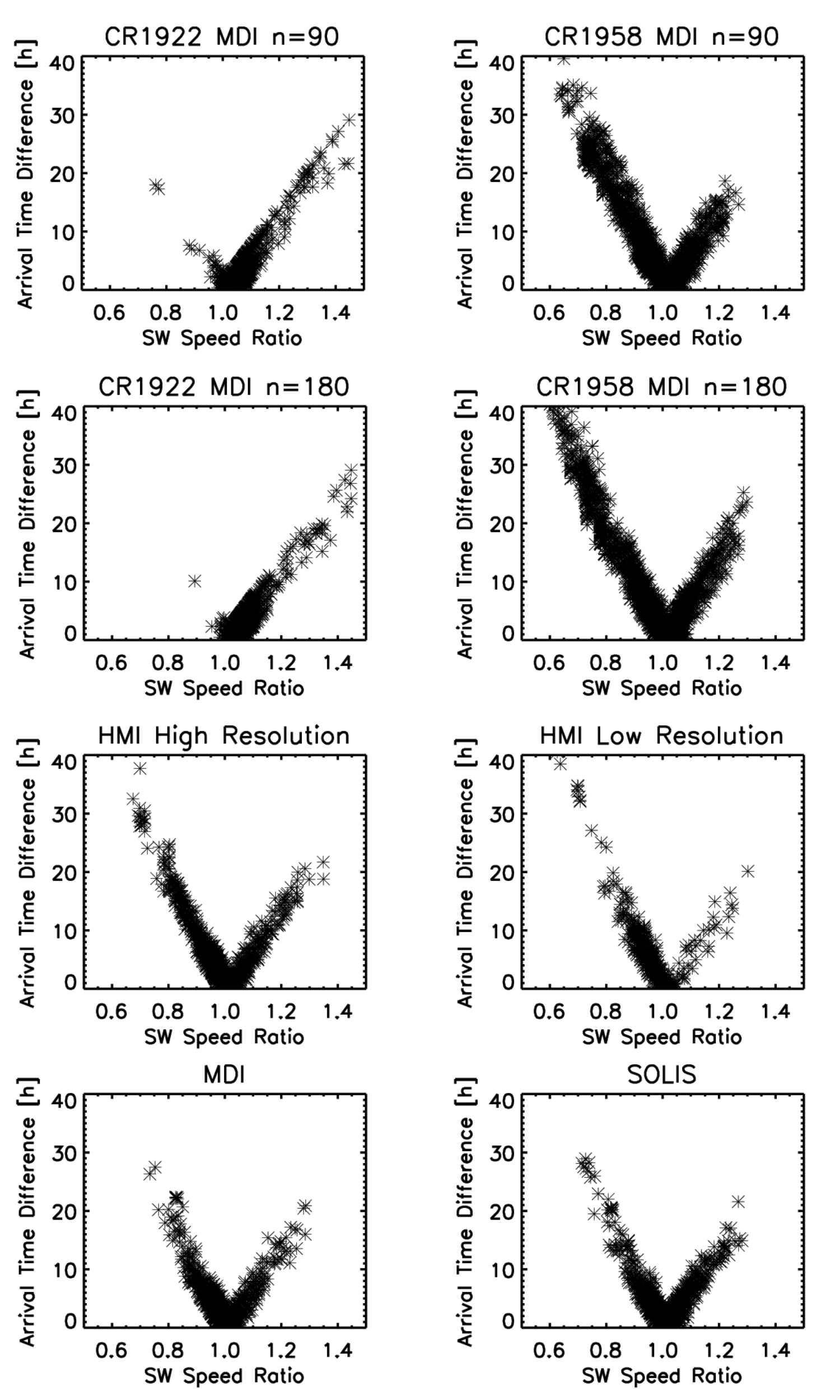}
\caption{Differences in wind arrival time to Earth as a function of the ratio between the wind speed calculated at the SS and at the AS ($u_\mathrm{ss}/u_\mathrm{as}$), based on field line tracing.}
\label{fig:f8}
\end{figure}

\subsection{Impact on the Solar Wind Arrival Time to Earth}
\label{ArrivalTime}

Beyond its physical interpretation, the relation between the expansion factor and the solar-wind speed is used by the WSA model to predict the arrival time of a solar wind element to the Earth. We use simple ballistic propagation of the solar-wind speed obtained from Equation~\ref{Eq:7} to calculate the solar-wind travel time from the SS, and from the AS to 1~AU. Figure~\ref{fig:f8} shows the arrival-time difference (in hours) as a function of the ratio $u_\mathrm{ss}/u_\mathrm{as}$ for a given magnetic field line. Most of the points are located between 10\,\% or less speed difference (ratio between 0.9 to 1.1), and an arrival-time difference of about five hours or less. However, a significant fraction of points have speed difference of 20\,\% or more (ratio of less than 0.8 or more than 1.2), and an arrival-time difference of more than ten hours. The fractions of points with a speed ratio above 10\,\% and above 20\,\% for all cases are summarized in Teble~\ref{Table:t5}. It shows that for solar minimum and fast-solar-wind speeds, the arrival time difference is more than five hours for about 20\,\% of the field lines, and for about 40\,\% (nearly half) of the field lines for solar maximum and slow solar-wind speeds. The solutions for CR 2104 seems to represent an intermittent stage.

\begin{figure}[h!]
\centering
\noindent\includegraphics[width=4.75in]{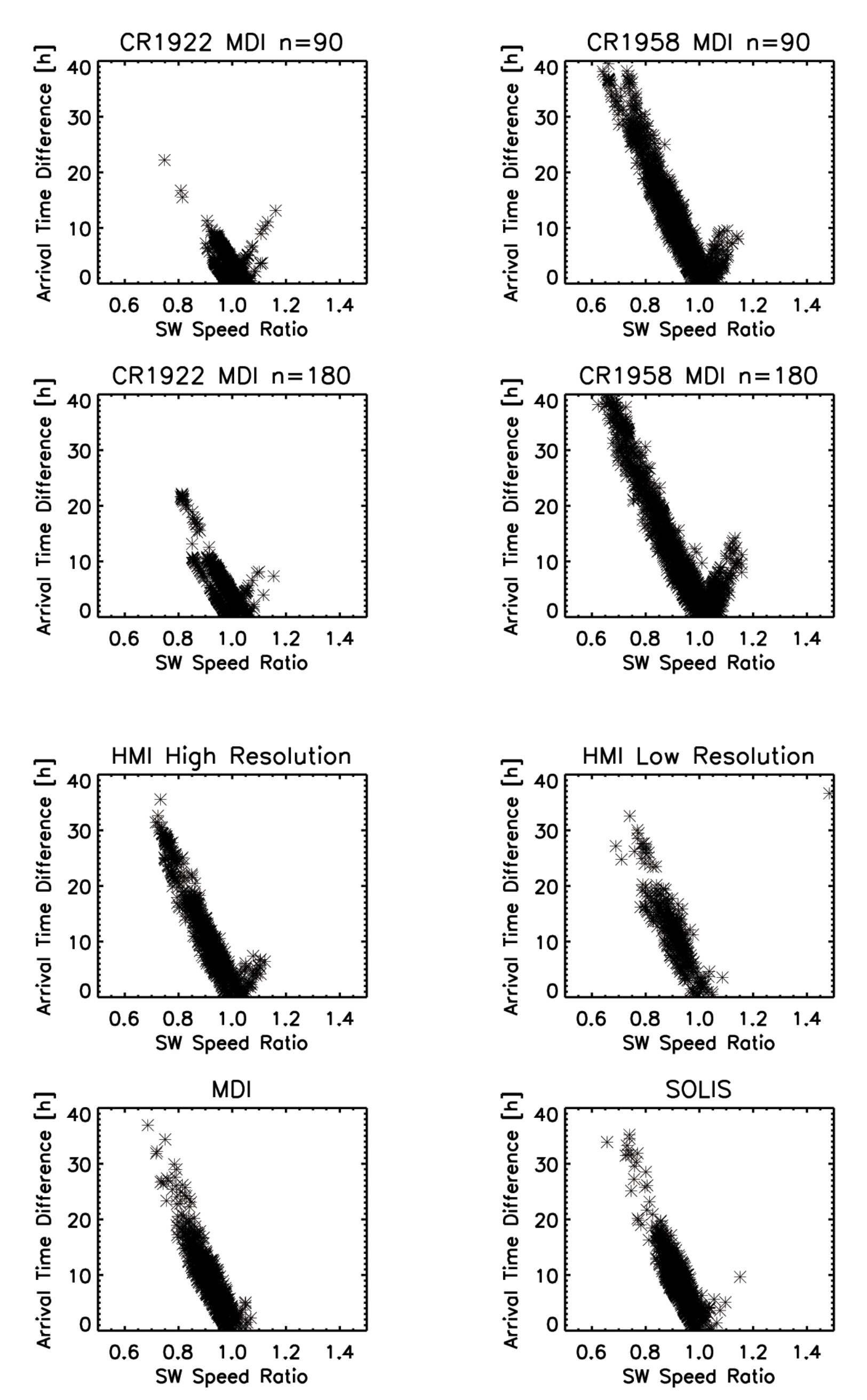}
\caption{Differences in wind arrival time to Earth as a function of the ratio between the wind speed calculated at the SS and at the AS ($u_\mathrm{ss}/u_\mathrm{as}$), based on radial propagation.}
\label{fig:f9}
\end{figure}

Figure~\ref{fig:f9} shows a similar calculation for the radial propagation from the SS to the AS. Here, the difference in the arrival time for CR 1922 becomes very small comparing to these cases with the full field line tracing. For the other cases, the solar-wind speed ratio appears mostly for the values of less than one, in contrast to the symmetric structure in Figure~\ref{fig:f8}. This is due to the consistency of $f_\mathrm{s}(SS)>f_\mathrm{s}(AS)$.

\begin{table}
\caption{Fraction of field lines with solar-wind speed ratio above 10\,\% and 20\,\% for field line tracing from the SS to the AS}
\begin{adjustbox}{max width=4in}
\begin{tabular}{ccc}
\hline
Magnetogram & solar-wind speed &  solar-wind speed \\
Input  & Ratio Above 10\,\% &  Above 20\,\% \\
\hline
MDI CR 1922 n=90 &  18\,\% & 6\,\%\\
MDI CR 1922 n=180 & 20\,\%& 5\,\%\\
MDI CR 1958 n=90 & 35\,\% & 16\,\%\\
MDI CR 1958 n=180 & 43\,\% & 26\,\%\\
HMI high-res CR 2104 & 26\,\% & 9\,\%\\
HMI low-res CR 2104 & 26\,\% & 6\,\%\\
MDI CR 2104 & 19\,\% & 4\,\%\\
SOLIS CR 2104 & 20\,\% & 6\,\%\\
\hline
\end{tabular}
\end{adjustbox}
\label{Table:t5}
\end{table}

\begin{table}
\caption{Fraction of field lines with solar-wind speed ratio above 10\,\% and 20\,\% for radial propagation from the SS to the AS}
\begin{adjustbox}{max width=4in}
\begin{tabular}{ccc}
\hline
Magnetogram & solar-wind speed &  solar-wind speed \\
Input  & Ratio Above 10\,\% &  Above 20\,\% \\
\hline
MDI CR 1922 n=90 &  2\,\% & 1\,\%\\
MDI CR 1922 n=180 & 45\,\%& 2\,\%\\
MDI CR 1958 n=90 & 9\,\% & 2\,\%\\
MDI CR 1958 n=180 & 46\,\% & 27\,\%\\
HMI high-res CR 2104 & 39\,\% & 15\,\%\\
HMI low-res CR 2104 & 48\,\% & 12\,\%\\
MDI CR 2104 & 41\,\% & 6\,\%\\
SOLIS CR 2104 & 41\,\% & 3\,\%\\
\hline
\end{tabular}
\end{adjustbox}
\label{Table:t6}
\end{table}

\section{Summary and Conclusions}
\label{Conclusions}

In this article, we investigate the difference in the field geometry between a global MHD model and the potential field source surface model (PFSSM) by tracing individual magnetic field lines from the Alfv\'en surface (AS), through the source surface (SS), all the way to the field line footpoint on the solar surface, and then trace them back to the SS in the PFSSM. We also compare two points along the same radial line at the SS and the AS. We study the effect of solar-cycle variations, the order of the harmonic expansion, and the magnetogram source. 

We find that for solar-minimum periods, the solutions are associated with small values of $f_\mathrm{s}$ and fast solar-wind speeds, and the PFSSM under estimates the flux-tube expansion due to the further change of the filed geometry beyond the SS. For the solar-maximum case, the solutions are associated with large values of $f_\mathrm{s}$ and slow solar-wind speeds. Here, we find that the PFSSM over-estimates the expansion factor due to the truncation of the magnetic loops and the radial forcing of the field at the SS, which leads to a consistent change in the field geometry between the potential field solution and the steady-state MHD solution.

Finally, we estimate the difference in the solar-wind arrival time to 1~AU between the two models. We find that the deviation of the arrival-time between the two models for 20\,\% of the field lines in the solar-minimum case, and for 40\,\% of the field lines in the solar-maximum case is more than five hours.

\section* {Disclosure of Potential Conflicts of Interest}

The author, Ofer Cohen declares that he has no conflicts of interest.

%
\begin{acks}
Simulation results were obtained using the Space Weather Modeling Framework, developed by the Center for Space Environment Modeling, at the University of Michigan with funding support from NASA ESS, NASA ESTO-CT, NSF KDI, and DoD MURI.
 \end{acks}



\begin{thebibliography}{34}
\ifx\bisbn     \undefined \def\bisbn  #1{ISBN #1}\fi
\ifx\binits    \undefined \def\binits#1{#1}\fi
\ifx\bauthor   \undefined \def\bauthor#1{#1}\fi
\ifx\batitle   \undefined \def\batitle#1{#1}\fi
\ifx\bjtitle   \undefined \def\bjtitle#1{\textit{#1}}\fi
\ifx\bvolume   \undefined \def\bvolume#1{\textbf{#1}}\fi
\ifx\byear     \undefined \def\byear#1{#1}\fi
\ifx\bissue    \undefined \def\bissue#1{#1}\fi
\ifx\bfpage    \undefined \def\bfpage#1{#1}\fi
\ifx\blpage    \undefined \def\blpage #1{#1}\fi
\ifx\burl      \undefined \def\burl#1{\textsf{#1}}\fi
\ifx\href      \undefined \def\href#1#2{\textsf{#2}}\fi
\ifx\betal     \undefined \def\betal{\textit{et al.}}\fi
\ifx\bctitle   \undefined \def\bctitle#1{#1}\fi
\ifx\beditor   \undefined \def\beditor#1{#1}\fi
\ifx\bbtitle   \undefined \def\bbtitle#1{\textit{#1}}\fi
\ifx\bedition  \undefined \def\bedition#1{#1}\fi
\ifx\bseriesno \undefined \def\bseriesno#1{\textbf{#1}}\fi
\ifx\blocation \undefined \def\blocation#1{#1}\fi
\ifx\bsertitle \undefined \def\bsertitle#1{\textit{#1}}\fi
\ifx\bsnm      \undefined \def\bsnm#1{#1}\fi
\ifx\bsuffix   \undefined \def\bsuffix#1{#1}\fi
\ifx\bparticle \undefined \def\bparticle#1{#1}\fi
\ifx\barticle  \undefined \def\barticle#1{}\fi
\ifx\binstitute  \undefined \def\binstitute#1{#1}\fi
\ifx\bpublisher  \undefined \def\bpublisher#1{#1}\fi
\ifx\doiurl    \undefined
  \def\doiurl#1{\href{http://dx.doi.org/#1}{\textsf{DOI}}}\fi
\ifx\arxivurl  \undefined
  \def\arxivurl#1{\href{http://arxiv.org/abs/#1}{\textsf{arXiv}}}\fi
\ifx\adsurl    \undefined
  \def\adsurl#1{\href{http://adsabs.harvard.edu/abs/#1}{\textsf{ADS}}}\fi
\ifx\botherref \undefined \def\botherref#1{}\fi
\ifx\url       \undefined \def\url#1{\textsf{#1}}\fi
\ifx\bchapter  \undefined \def\bchapter#1{}\fi
\ifx\bbook     \undefined \def\bbook#1{}\fi
\ifx\bcomment  \undefined \def\bcomment#1{#1}\fi
\ifx\oauthor   \undefined \def\oauthor#1{#1}\fi
\ifx\citeauthoryear \undefined\def \citeauthoryear#1{#1}\fi
\def\endbibitem {}
\ifx\bconflocation  \undefined \def\bconflocation#1{#1} \fi

\bibitem[\protect\citeauthoryear{{Altschuler} and
  {Newkirk}}{1969}]{AltschulerNewkirk69}
\begin{barticle}
\bauthor{\bsnm{{Altschuler}}, \binits{M.D.}},
\bauthor{\bsnm{{Newkirk}}, \binits{G.}}:
\byear{1969},
\batitle{{Magnetic Fields and the Structure of the Solar Corona. I: Methods of
  Calculating Coronal Fields}}.
\bjtitle{Solar Phys.}
\bvolume{9},
\bfpage{131}.
\doiurl{10.1007/BF00145734}.
\adsurl{1969SoPh....9..131A}.
\end{barticle}
\endbibitem

\bibitem[\protect\citeauthoryear{{Altschuler}
  \textit{et~al.}}{1977}]{Altschuler77}
\begin{barticle}
\bauthor{\bsnm{{Altschuler}}, \binits{M.D.}},
\bauthor{\bsnm{{Levine}}, \binits{R.H.}},
\bauthor{\bsnm{{Stix}}, \binits{M.}},
\bauthor{\bsnm{{Harvey}}, \binits{J.}}:
\byear{1977},
\batitle{{High Resolution Mapping of the Magnetic Field of the Solar Corona}}.
\bjtitle{Solar Phys.}
\bvolume{51},
\bfpage{345}.
\adsurl{cgi-bin/nph-bib_query?bibcode=1977SoPh...51..345A&db_key=AST}.
\end{barticle}
\endbibitem

\bibitem[\protect\citeauthoryear{{Arge} and {Pizzo}}{2000}]{ArgePizzo00}
\begin{barticle}
\bauthor{\bsnm{{Arge}}, \binits{C.N.}},
\bauthor{\bsnm{{Pizzo}}, \binits{V.J.}}:
\byear{2000},
\batitle{{Improvement in the Prediction of Solar Wind Conditions Using
  Near-Real time Solar Magnetic Field Updates}}.
\bjtitle{Journal of Geophysical Research (Space Physics)}
\bvolume{105},
\bfpage{10,465}.
\doiurl{10.1029/1999JA900262}.
\adsurl{cgi-bin/nph-bib_query?bibcode=2000JGR...105.7509C&db_key=AST}.
\end{barticle}
\endbibitem

\bibitem[\protect\citeauthoryear{{Cohen} \textit{et~al.}}{2007}]{Cohen07}
\begin{barticle}
\bauthor{\bsnm{{Cohen}}, \binits{O.}},
\bauthor{\bsnm{{Sokolov}}, \binits{I.V.}},
\bauthor{\bsnm{{Roussev}}, \binits{I.I.}},
\bauthor{\bsnm{{Arge}}, \binits{C.N.}},
\bauthor{\bsnm{{Manchester}}, \binits{W.B.}},
\bauthor{\bsnm{{Gombosi}}, \binits{T.I.}},
\bauthor{\bsnm{{Frazin}}, \binits{R.A.}},
\bauthor{\bsnm{{Park}}, \binits{H.}},
\bauthor{\bsnm{{Butala}}, \binits{M.D.}},
\bauthor{\bsnm{{Kamalabadi}}, \binits{F.}},
\bauthor{\bsnm{{Velli}}, \binits{M.}}:
\byear{2007},
\batitle{{A Semiempirical Magnetohydrodynamical Model of the Solar Wind}}.
\bjtitle{Astrophys. J. Lett.}
\bvolume{654},
\bfpage{L163}.
\doiurl{10.1086/511154}.
\adsurl{2007ApJ...654L.163C}.
\end{barticle}
\endbibitem

\bibitem[\protect\citeauthoryear{{DeForest}, {Howard}, and
  {McComas}}{2014}]{DeForest13}
\begin{barticle}
\bauthor{\bsnm{{DeForest}}, \binits{C.E.}},
\bauthor{\bsnm{{Howard}}, \binits{T.A.}},
\bauthor{\bsnm{{McComas}}, \binits{D.J.}}:
\byear{2014},
\batitle{{Inbound Waves in the Solar Corona: A Direct Indicator of Alfv{\'e}n
  Surface Location}}.
\bjtitle{Astrophys. J.}
\bvolume{787},
\bfpage{124}.
\doiurl{10.1088/0004-637X/787/2/124}.
\adsurl{2014ApJ...787..124D}.
\end{barticle}
\endbibitem

\bibitem[\protect\citeauthoryear{{Gilbert}, {Zurbuchen}, and
  {Fisk}}{2007}]{Gilbert07}
\begin{barticle}
\bauthor{\bsnm{{Gilbert}}, \binits{J.A.}},
\bauthor{\bsnm{{Zurbuchen}}, \binits{T.H.}},
\bauthor{\bsnm{{Fisk}}, \binits{L.A.}}:
\byear{2007},
\batitle{{A New Technique for Mapping Open Magnetic Flux from the Solar Surface
  into the Heliosphere}}.
\bjtitle{Astrophys. J.}
\bvolume{663},
\bfpage{583}.
\doiurl{10.1086/518099}.
\adsurl{2007ApJ...663..583G}.
\end{barticle}
\endbibitem

\bibitem[\protect\citeauthoryear{{Hoeksema}}{1984}]{Hoeksema84}
\begin{botherref}
\oauthor{\bsnm{{Hoeksema}}, \binits{J.T.}}:
1984,
{Structure and evolution of the large scale solar and heliospheric magnetic
  fields}.
PhD thesis,
Stanford Univ., CA.
\adsurl{1984PhDT.........5H}.
\end{botherref}
\endbibitem

\bibitem[\protect\citeauthoryear{{Lee} \textit{et~al.}}{2009}]{Lee08}
\begin{barticle}
\bauthor{\bsnm{{Lee}}, \binits{C.O.}},
\bauthor{\bsnm{{Luhmann}}, \binits{J.G.}},
\bauthor{\bsnm{{Odstrcil}}, \binits{D.}},
\bauthor{\bsnm{{MacNeice}}, \binits{P.J.}},
\bauthor{\bsnm{{de Pater}}, \binits{I.}},
\bauthor{\bsnm{{Riley}}, \binits{P.}},
\bauthor{\bsnm{{Arge}}, \binits{C.N.}}:
\byear{2009},
\batitle{{The Solar Wind at 1 AU During the Declining Phase of Solar Cycle 23:
  Comparison of 3D Numerical Model Results with Observations}}.
\bjtitle{Solar Phys.}
\bvolume{254},
\bfpage{155}.
\doiurl{10.1007/s11207-008-9280-y}.
\adsurl{2009SoPh..254..155L}.
\end{barticle}
\endbibitem

\bibitem[\protect\citeauthoryear{{Lee} \textit{et~al.}}{2011}]{Lee11}
\begin{barticle}
\bauthor{\bsnm{{Lee}}, \binits{C.O.}},
\bauthor{\bsnm{{Luhmann}}, \binits{J.G.}},
\bauthor{\bsnm{{Hoeksema}}, \binits{J.T.}},
\bauthor{\bsnm{{Sun}}, \binits{X.}},
\bauthor{\bsnm{{Arge}}, \binits{C.N.}},
\bauthor{\bsnm{{de Pater}}, \binits{I.}}:
\byear{2011},
\batitle{{Coronal Field Opens at Lower Height During the Solar Cycles 22 and 23
  Minimum Periods: IMF Comparison Suggests the Source Surface Should Be
  Lowered}}.
\bjtitle{Solar Phys.}
\bvolume{269},
\bfpage{367}.
\doiurl{10.1007/s11207-010-9699-9}.
\adsurl{2011SoPh..269..367L}.
\end{barticle}
\endbibitem

\bibitem[\protect\citeauthoryear{{Levine}, {Schulz}, and
  {Frazier}}{1982}]{Levine82}
\begin{barticle}
\bauthor{\bsnm{{Levine}}, \binits{R.H.}},
\bauthor{\bsnm{{Schulz}}, \binits{M.}},
\bauthor{\bsnm{{Frazier}}, \binits{E.N.}}:
\byear{1982},
\batitle{{Simulation of the magnetic structure of the inner heliosphere by
  means of a non-spherical source surface}}.
\bjtitle{Solar Phys.}
\bvolume{77},
\bfpage{363}.
\doiurl{10.1007/BF00156118}.
\adsurl{1982SoPh...77..363L}.
\end{barticle}
\endbibitem

\bibitem[\protect\citeauthoryear{{Linker} \textit{et~al.}}{1999}]{Linker99}
\begin{barticle}
\bauthor{\bsnm{{Linker}}, \binits{J.A.}},
\bauthor{\bsnm{{Miki{\'c}}}, \binits{Z.}},
\bauthor{\bsnm{{Biesecker}}, \binits{D.A.}},
\bauthor{\bsnm{{Forsyth}}, \binits{R.J.}},
\bauthor{\bsnm{{Gibson}}, \binits{S.E.}},
\bauthor{\bsnm{{Lazarus}}, \binits{A.J.}},
\bauthor{\bsnm{{Lecinski}}, \binits{A.}},
\bauthor{\bsnm{{Riley}}, \binits{P.}},
\bauthor{\bsnm{{Szabo}}, \binits{A.}},
\bauthor{\bsnm{{Thompson}}, \binits{B.J.}}:
\byear{1999},
\batitle{{Magnetohydrodynamic modeling of the solar corona during Whole Sun
  Month}}.
\bjtitle{Journal of Geophysical Research (Space Physics)}
\bvolume{104},
\bfpage{9809}.
\doiurl{10.1029/1998JA900159}.
\adsurl{1999JGR...104.9809L}.
\end{barticle}
\endbibitem

\bibitem[\protect\citeauthoryear{{Lionello}, {Linker}, and
  {Miki{\'c}}}{2009}]{Lionello09}
\begin{barticle}
\bauthor{\bsnm{{Lionello}}, \binits{R.}},
\bauthor{\bsnm{{Linker}}, \binits{J.A.}},
\bauthor{\bsnm{{Miki{\'c}}}, \binits{Z.}}:
\byear{2009},
\batitle{{Multispectral Emission of the Sun During the First Whole Sun Month:
  Magnetohydrodynamic Simulations}}.
\bjtitle{Astrophys. J.}
\bvolume{690},
\bfpage{902}.
\doiurl{10.1088/0004-637X/690/1/902}.
\adsurl{2009ApJ...690..902L}.
\end{barticle}
\endbibitem

\bibitem[\protect\citeauthoryear{{Lionello}, {Mikic}, and
  {Schnack}}{1998}]{Lionello98}
\begin{barticle}
\bauthor{\bsnm{{Lionello}}, \binits{R.}},
\bauthor{\bsnm{{Mikic}}, \binits{Z.}},
\bauthor{\bsnm{{Schnack}}, \binits{D.D.}}:
\byear{1998},
\batitle{{Magnetohydrodynamics of solar coronal plasmas in cylindrical
  geometry.}}
\bjtitle{Journal of Computational Physics}
\bvolume{140},
\bfpage{172}.
\doiurl{10.1006/jcph.1998.5841}.
\adsurl{1998JCoPh.140..172L}.
\end{barticle}
\endbibitem

\bibitem[\protect\citeauthoryear{{Luhmann} \textit{et~al.}}{2002}]{Luhmann02}
\begin{barticle}
\bauthor{\bsnm{{Luhmann}}, \binits{J.G.}},
\bauthor{\bsnm{{Li}}, \binits{Y.}},
\bauthor{\bsnm{{Arge}}, \binits{C.N.}},
\bauthor{\bsnm{{Gazis}}, \binits{P.R.}},
\bauthor{\bsnm{{Ulrich}}, \binits{R.}}:
\byear{2002},
\batitle{{Solar cycle changes in coronal holes and space weather cycles}}.
\bjtitle{Journal of Geophysical Research (Space Physics)}
\bvolume{107},
\bfpage{1154}.
\doiurl{10.1029/2001JA007550}.
\adsurl{2002JGRA..107.1154L}.
\end{barticle}
\endbibitem

\bibitem[\protect\citeauthoryear{{McComas} \textit{et~al.}}{2007}]{McComas07}
\begin{barticle}
\bauthor{\bsnm{{McComas}}, \binits{D.J.}},
\bauthor{\bsnm{{Velli}}, \binits{M.}},
\bauthor{\bsnm{{Lewis}}, \binits{W.S.}},
\bauthor{\bsnm{{Acton}}, \binits{L.W.}},
\bauthor{\bsnm{{Balat-Pichelin}}, \binits{M.}},
\bauthor{\bsnm{{Bothmer}}, \binits{V.}},
\bauthor{\bsnm{{Dirling}}, \binits{R.B.}},
\bauthor{\bsnm{{Feldman}}, \binits{W.C.}},
\bauthor{\bsnm{{Gloeckler}}, \binits{G.}},
\bauthor{\bsnm{{Habbal}}, \binits{S.R.}},
\bauthor{\bsnm{{Hassler}}, \binits{D.M.}},
\bauthor{\bsnm{{Mann}}, \binits{I.}},
\bauthor{\bsnm{{Matthaeus}}, \binits{W.H.}},
\bauthor{\bsnm{{McNutt}}, \binits{R.L.}},
\bauthor{\bsnm{{Mewaldt}}, \binits{R.A.}},
\bauthor{\bsnm{{Murphy}}, \binits{N.}},
\bauthor{\bsnm{{Ofman}}, \binits{L.}},
\bauthor{\bsnm{{Sittler}}, \binits{E.C.}},
\bauthor{\bsnm{{Smith}}, \binits{C.W.}},
\bauthor{\bsnm{{Zurbuchen}}, \binits{T.H.}}:
\byear{2007},
\batitle{{Understanding coronal heating and solar wind acceleration: Case for
  in situ near-Sun measurements}}.
\bjtitle{Reviews of Geophysics}
\bvolume{45},
\bfpage{1004}.
\doiurl{10.1029/2006RG000195}.
\adsurl{2007RvGeo..45.1004M}.
\end{barticle}
\endbibitem

\bibitem[\protect\citeauthoryear{{McGregor} \textit{et~al.}}{2011}]{McGregor11}
\begin{barticle}
\bauthor{\bsnm{{McGregor}}, \binits{S.L.}},
\bauthor{\bsnm{{Hughes}}, \binits{W.J.}},
\bauthor{\bsnm{{Arge}}, \binits{C.N.}},
\bauthor{\bsnm{{Owens}}, \binits{M.J.}},
\bauthor{\bsnm{{Odstrcil}}, \binits{D.}}:
\byear{2011},
\batitle{{The distribution of solar-wind speeds during solar minimum:
  Calibration for numerical solar wind modeling constraints on the source of
  the slow solar wind}}.
\bjtitle{Journal of Geophysical Research (Space Physics)}
\bvolume{116},
\bfpage{3101}.
\doiurl{10.1029/2010JA015881}.
\adsurl{2011JGRA..116.3101M}.
\end{barticle}
\endbibitem

\bibitem[\protect\citeauthoryear{{Mikic} and {Linker}}{1994}]{MikicLinker94}
\begin{barticle}
\bauthor{\bsnm{{Mikic}}, \binits{Z.}},
\bauthor{\bsnm{{Linker}}, \binits{J.A.}}:
\byear{1994},
\batitle{{Disruption of coronal magnetic field arcades}}.
\bjtitle{Astrophys. J.}
\bvolume{430},
\bfpage{898}.
\doiurl{10.1086/174460}.
\adsurl{1994ApJ...430..898M}.
\end{barticle}
\endbibitem

\bibitem[\protect\citeauthoryear{{Miki{\'c}} \textit{et~al.}}{1999}]{Mikic99}
\begin{barticle}
\bauthor{\bsnm{{Miki{\'c}}}, \binits{Z.}},
\bauthor{\bsnm{{Linker}}, \binits{J.A.}},
\bauthor{\bsnm{{Schnack}}, \binits{D.D.}},
\bauthor{\bsnm{{Lionello}}, \binits{R.}},
\bauthor{\bsnm{{Tarditi}}, \binits{A.}}:
\byear{1999},
\batitle{{Magnetohydrodynamic modeling of the global solar corona}}.
\bjtitle{Phys. Plasmas}
\bvolume{6},
\bfpage{2217}.
\doiurl{10.1063/1.873474}.
\adsurl{1999PhPl....6.2217M}.
\end{barticle}
\endbibitem

\bibitem[\protect\citeauthoryear{{Odstrcil}}{2003}]{Odstrcil03}
\begin{barticle}
\bauthor{\bsnm{{Odstrcil}}, \binits{D.}}:
\byear{2003},
\batitle{{Modeling 3-D solar wind structure}}.
\bjtitle{Advances in Space Research}
\bvolume{32},
\bfpage{497}.
\doiurl{10.1016/S0273-1177(03)00332-6}.
\adsurl{2003AdSpR..32..497O}.
\end{barticle}
\endbibitem

\bibitem[\protect\citeauthoryear{{Oran} \textit{et~al.}}{2013}]{Oran13}
\begin{barticle}
\bauthor{\bsnm{{Oran}}, \binits{R.}},
\bauthor{\bsnm{{van der Holst}}, \binits{B.}},
\bauthor{\bsnm{{Landi}}, \binits{E.}},
\bauthor{\bsnm{{Jin}}, \binits{M.}},
\bauthor{\bsnm{{Sokolov}}, \binits{I.V.}},
\bauthor{\bsnm{{Gombosi}}, \binits{T.I.}}:
\byear{2013},
\batitle{{A Global Wave-driven Magnetohydrodynamic Solar Model with a Unified
  Treatment of Open and Closed Magnetic Field Topologies}}.
\bjtitle{Astrophys. J.}
\bvolume{778},
\bfpage{176}.
\doiurl{10.1088/0004-637X/778/2/176}.
\adsurl{2013ApJ...778..176O}.
\end{barticle}
\endbibitem

\bibitem[\protect\citeauthoryear{{Pneuman} and {Kopp}}{1971}]{PneumanKopp71}
\begin{barticle}
\bauthor{\bsnm{{Pneuman}}, \binits{G.W.}},
\bauthor{\bsnm{{Kopp}}, \binits{R.A.}}:
\byear{1971},
\batitle{{Gas-Magnetic Field Interactions in the Solar Corona}}.
\bjtitle{Solar Phys.}
\bvolume{18},
\bfpage{258}.
\doiurl{10.1007/BF00145940}.
\adsurl{1971SoPh...18..258P}.
\end{barticle}
\endbibitem

\bibitem[\protect\citeauthoryear{{Poduval} and {Zhao}}{2014}]{PoduvalZhao14}
\begin{barticle}
\bauthor{\bsnm{{Poduval}}, \binits{B.}},
\bauthor{\bsnm{{Zhao}}, \binits{X.P.}}:
\byear{2014},
\batitle{{Validating Solar Wind Prediction Using the Current Sheet Source
  Surface Model}}.
\bjtitle{Astrophys. J. Lett.}
\bvolume{782},
\bfpage{L22}.
\doiurl{10.1088/2041-8205/782/2/L22}.
\adsurl{2014ApJ...782L..22P}.
\end{barticle}
\endbibitem

\bibitem[\protect\citeauthoryear{{Powell} \textit{et~al.}}{1999}]{Powell99}
\begin{barticle}
\bauthor{\bsnm{{Powell}}, \binits{K.G.}},
\bauthor{\bsnm{{Roe}}, \binits{P.L.}},
\bauthor{\bsnm{{Linde}}, \binits{T.J.}},
\bauthor{\bsnm{{Gombosi}}, \binits{T.I.}},
\bauthor{\bsnm{{de Zeeuw}}, \binits{D.L.}}:
\byear{1999},
\batitle{{A Solution-Adaptive Upwind Scheme for Ideal Magnetohydrodynamics}}.
\bjtitle{Journal of Computational Physics}
\bvolume{154},
\bfpage{284}.
\doiurl{10.1006/jcph.1999.6299}.
\adsurl{1999JCoPh.154..284P}.
\end{barticle}
\endbibitem

\bibitem[\protect\citeauthoryear{{Riley} \textit{et~al.}}{2006}]{Riley06}
\begin{barticle}
\bauthor{\bsnm{{Riley}}, \binits{P.}},
\bauthor{\bsnm{{Linker}}, \binits{J.A.}},
\bauthor{\bsnm{{Miki{\'c}}}, \binits{Z.}},
\bauthor{\bsnm{{Lionello}}, \binits{R.}},
\bauthor{\bsnm{{Ledvina}}, \binits{S.A.}},
\bauthor{\bsnm{{Luhmann}}, \binits{J.G.}}:
\byear{2006},
\batitle{{A Comparison between Global Solar Magnetohydrodynamic and Potential
  Field Source Surface Model Results}}.
\bjtitle{Astrophys. J.}
\bvolume{653},
\bfpage{1510}.
\doiurl{10.1086/508565}.
\adsurl{2006ApJ...653.1510R}.
\end{barticle}
\endbibitem

\bibitem[\protect\citeauthoryear{{Riley} \textit{et~al.}}{2011}]{Riley11}
\begin{barticle}
\bauthor{\bsnm{{Riley}}, \binits{P.}},
\bauthor{\bsnm{{Lionello}}, \binits{R.}},
\bauthor{\bsnm{{Linker}}, \binits{J.A.}},
\bauthor{\bsnm{{Mikic}}, \binits{Z.}},
\bauthor{\bsnm{{Luhmann}}, \binits{J.}},
\bauthor{\bsnm{{Wijaya}}, \binits{J.}}:
\byear{2011},
\batitle{{Global MHD Modeling of the Solar Corona and Inner Heliosphere for the
  Whole Heliosphere Interval}}.
\bjtitle{Solar Phys.}
\bvolume{274},
\bfpage{361}.
\doiurl{10.1007/s11207-010-9698-x}.
\adsurl{2011SoPh..274..361R}.
\end{barticle}
\endbibitem

\bibitem[\protect\citeauthoryear{{Ru{\v s}in} \textit{et~al.}}{2010}]{Rusin10}
\begin{barticle}
\bauthor{\bsnm{{Ru{\v s}in}}, \binits{V.}},
\bauthor{\bsnm{{Druckm{\"u}ller}}, \binits{M.}},
\bauthor{\bsnm{{Aniol}}, \binits{P.}},
\bauthor{\bsnm{{Minarovjech}}, \binits{M.}},
\bauthor{\bsnm{{Saniga}}, \binits{M.}},
\bauthor{\bsnm{{Miki{\'c}}}, \binits{Z.}},
\bauthor{\bsnm{{Linker}}, \binits{J.A.}},
\bauthor{\bsnm{{Lionello}}, \binits{R.}},
\bauthor{\bsnm{{Riley}}, \binits{P.}},
\bauthor{\bsnm{{Titov}}, \binits{V.S.}}:
\byear{2010},
\batitle{{Comparing eclipse observations of the 2008 August 1 solar corona with
  an MHD model prediction}}.
\bjtitle{A\&A}
\bvolume{513},
\bfpage{A45}.
\doiurl{10.1051/0004-6361/200912778}.
\adsurl{2010A\,\%26A...513A..45R}.
\end{barticle}
\endbibitem

\bibitem[\protect\citeauthoryear{{Schulz}}{1997}]{Schulz97}
\begin{barticle}
\bauthor{\bsnm{{Schulz}}, \binits{M.}}:
\byear{1997},
\batitle{{Non-spherical source-surface model of the heliosphere: a scalar
  formulation}}.
\bjtitle{Annales Geophysicae}
\bvolume{15},
\bfpage{1379}.
\doiurl{10.1007/s00585-997-1379-1}.
\adsurl{1997AnGeo..15.1379S}.
\end{barticle}
\endbibitem

\bibitem[\protect\citeauthoryear{{Schulz}, {Frazier}, and
  {Boucher}}{1978}]{Schulz78}
\begin{barticle}
\bauthor{\bsnm{{Schulz}}, \binits{M.}},
\bauthor{\bsnm{{Frazier}}, \binits{E.N.}},
\bauthor{\bsnm{{Boucher}}, \binits{D.J.} \bsuffix{Jr.}}:
\byear{1978},
\batitle{{Coronal magnetic-field model with non-spherical source surface}}.
\bjtitle{Solar Phys.}
\bvolume{60},
\bfpage{83}.
\doiurl{10.1007/BF00152334}.
\adsurl{1978SoPh...60...83S}.
\end{barticle}
\endbibitem

\bibitem[\protect\citeauthoryear{{Sokolov} \textit{et~al.}}{2013}]{Sokolov13}
\begin{barticle}
\bauthor{\bsnm{{Sokolov}}, \binits{I.V.}},
\bauthor{\bsnm{{van der Holst}}, \binits{B.}},
\bauthor{\bsnm{{Oran}}, \binits{R.}},
\bauthor{\bsnm{{Downs}}, \binits{C.}},
\bauthor{\bsnm{{Roussev}}, \binits{I.I.}},
\bauthor{\bsnm{{Jin}}, \binits{M.}},
\bauthor{\bsnm{{Manchester}}, \binits{W.B.} \bsuffix{IV}},
\bauthor{\bsnm{{Evans}}, \binits{R.M.}},
\bauthor{\bsnm{{Gombosi}}, \binits{T.I.}}:
\byear{2013},
\batitle{{Magnetohydrodynamic Waves and Coronal Heating: Unifying Empirical and
  MHD Turbulence Models}}.
\bjtitle{Astrophys. J.}
\bvolume{764},
\bfpage{23}.
\doiurl{10.1088/0004-637X/764/1/23}.
\adsurl{2013ApJ...764...23S}.
\end{barticle}
\endbibitem

\bibitem[\protect\citeauthoryear{{Toth} \textit{et~al.}}{2005}]{Toth05}
\begin{barticle}
\bauthor{\bsnm{{Toth}}, \binits{G.}},
\bauthor{\bsnm{{Sokolov}}, \binits{I.V.}},
\bauthor{\bsnm{{Gombosi}}, \binits{T.I.}},
\bauthor{\bsnm{{Chesney}}, \binits{D.R.}},
\bauthor{\bsnm{{Clauer}}, \binits{C.R.}},
\bauthor{\bsnm{{De Zeeuw}}, \binits{D.L.}},
\bauthor{\bsnm{{Hansen}}, \binits{K.C.}},
\bauthor{\bsnm{{Kane}}, \binits{K.J.}},
\bauthor{\bsnm{{Manchester}}, \binits{W.B.}},
\bauthor{\bsnm{{Oehmke}}, \binits{R.C.}},
\bauthor{\bsnm{{Powell}}, \binits{K.G.}},
\bauthor{\bsnm{{Ridley}}, \binits{A.J.}},
\bauthor{\bsnm{{Roussev}}, \binits{I.I.}},
\bauthor{\bsnm{{Stout}}, \binits{Q.F.}},
\bauthor{\bsnm{{Volberg}}, \binits{O.}},
\bauthor{\bsnm{{Wolf}}, \binits{R.A.}},
\bauthor{\bsnm{{Sazykin}}, \binits{S.}},
\bauthor{\bsnm{{Chan}}, \binits{A.}},
\bauthor{\bsnm{{Yu}}, \binits{B.}},
\bauthor{\bsnm{{K{\'o}ta}}, \binits{J.}}:
\byear{2005},
\batitle{{Space Weather Modeling Framework: A New Tool for the Space Science
  Community}}.
\bjtitle{Journal of Geophysical Research (Space Physics)}
\bvolume{110},
\bfpage{12,226}.
\doiurl{10.1029/2005JA011126}.
\adsurl{cgi-bin/nph-bib_query?bibcode=2005JGRA..11012226T&db_key=AST}.
\end{barticle}
\endbibitem

\bibitem[\protect\citeauthoryear{{T{\'o}th} \textit{et~al.}}{2012}]{Toth12}
\begin{barticle}
\bauthor{\bsnm{{T{\'o}th}}, \binits{G.}},
\bauthor{\bsnm{{van der Holst}}, \binits{B.}},
\bauthor{\bsnm{{Sokolov}}, \binits{I.V.}},
\bauthor{\bsnm{{De Zeeuw}}, \binits{D.L.}},
\bauthor{\bsnm{{Gombosi}}, \binits{T.I.}},
\bauthor{\bsnm{{Fang}}, \binits{F.}},
\bauthor{\bsnm{{Manchester}}, \binits{W.B.}},
\bauthor{\bsnm{{Meng}}, \binits{X.}},
\bauthor{\bsnm{{Najib}}, \binits{D.}},
\bauthor{\bsnm{{Powell}}, \binits{K.G.}},
\bauthor{\bsnm{{Stout}}, \binits{Q.F.}},
\bauthor{\bsnm{{Glocer}}, \binits{A.}},
\bauthor{\bsnm{{Ma}}, \binits{Y.-J.}},
\bauthor{\bsnm{{Opher}}, \binits{M.}}:
\byear{2012},
\batitle{{Adaptive numerical algorithms in space weather modeling}}.
\bjtitle{Journal of Computational Physics}
\bvolume{231},
\bfpage{870}.
\doiurl{10.1016/j.jcp.2011.02.006}.
\adsurl{2012JCoPh.231..870T}.
\end{barticle}
\endbibitem

\bibitem[\protect\citeauthoryear{{van der Holst}
  \textit{et~al.}}{2014}]{Vanderholst14}
\begin{barticle}
\bauthor{\bsnm{{van der Holst}}, \binits{B.}},
\bauthor{\bsnm{{Sokolov}}, \binits{I.V.}},
\bauthor{\bsnm{{Meng}}, \binits{X.}},
\bauthor{\bsnm{{Jin}}, \binits{M.}},
\bauthor{\bsnm{{Manchester}}, \binits{W.B.} \bsuffix{IV}},
\bauthor{\bsnm{{T{\'o}th}}, \binits{G.}},
\bauthor{\bsnm{{Gombosi}}, \binits{T.I.}}:
\byear{2014},
\batitle{{Alfv{\'e}n Wave Solar Model (AWSoM): Coronal Heating}}.
\bjtitle{Astrophys. J.}
\bvolume{782},
\bfpage{81}.
\doiurl{10.1088/0004-637X/782/2/81}.
\adsurl{2014ApJ...782...81V}.
\end{barticle}
\endbibitem

\bibitem[\protect\citeauthoryear{{Wang} and {Sheeley}}{1990}]{WangSheeley90}
\begin{barticle}
\bauthor{\bsnm{{Wang}}, \binits{Y.-M.}},
\bauthor{\bsnm{{Sheeley}}, \binits{N.R.}}:
\byear{1990},
\batitle{{solar-wind speed and Coronal Flux-Tube Expansion}}.
\bjtitle{Astrophys. J.}
\bvolume{355},
\bfpage{726}.
\adsurl{cgi-bin/nph-bib_query?bibcode=1990ApJ...355..726W&amp;db_key=AST}.
\end{barticle}
\endbibitem

\bibitem[\protect\citeauthoryear{{Wiegelmann} and
  {Sakurai}}{2012}]{WiegelmannSakurai12}
\begin{barticle}
\bauthor{\bsnm{{Wiegelmann}}, \binits{T.}},
\bauthor{\bsnm{{Sakurai}}, \binits{T.}}:
\byear{2012},
\batitle{{Solar Force-free Magnetic Fields}}.
\bjtitle{Living Reviews in Solar Physics}
\bvolume{9},
\bfpage{5}.
\doiurl{10.12942/lrsp-2012-5}.
\adsurl{2012LRSP....9....5W}.
\end{barticle}
\endbibitem

\end{thebibliography}

\end{article} 

\end{document}